\documentclass[twocolumn]{aastex631}
\let\tablenum\relax
\usepackage{siunitx}
\usepackage{amssymb, amsmath}
\usepackage{color}
\usepackage{ulem}
\usepackage{float}
\newcommand{\red}[1]{\textcolor{black}{#1}}

\usepackage{comment} 
\usepackage{booktabs}
\submitjournal{ApJ}
\begin{document}

\title{
Chemical Abundance Ratios of Nitrogen Rich Galaxies Identified at $z\sim 6-12$:\\
Observational Demographics and Models
}

\author[0000-0002-2740-3403]{Kuria Watanabe}
\affiliation{Department of Astronomical Science, SOKENDAI (The Graduate University for Advanced Studies), \\ 2-21-1 Osawa, Mitaka, Tokyo, 181-8588, Japan}
\affiliation{National Astronomical Observatory of Japan, 2-21-1 Osawa, Mitaka, Tokyo, 181-8588, Japan}

\author[0000-0002-1049-6658]{Masami Ouchi}
\affiliation{Department of Astronomical Science, SOKENDAI (The Graduate University for Advanced Studies), \\ 2-21-1 Osawa, Mitaka, Tokyo, 181-8588, Japan}
\affiliation{National Astronomical Observatory of Japan, 2-21-1 Osawa, Mitaka, Tokyo, 181-8588, Japan}
\affiliation{Institute for Cosmic Ray Research, The University of Tokyo, 5-1-5 Kashiwa-no-Ha, Kashiwa, Chiba, 277-8582, Japan}
\affiliation{Kavli Institute for the Physics and Mathematics of the Universe (WPI), The University of Tokyo, Kashiwa, Chiba 277-8583, Japan}

\author[0000-0003-2965-5070]{Kimihiko Nakajima}
\affiliation{Institute of Liberal Arts and Science, Kanazawa University, Kakuma-machi, Kanazawa, 920-1192, Ishikawa, Japan.}

\author[0000-0001-8537-3153]{Nozomu Tominaga}
\affiliation{Department of Astronomical Science, SOKENDAI (The Graduate University for Advanced Studies), \\ 2-21-1 Osawa, Mitaka, Tokyo, 181-8588, Japan}
\affiliation{National Astronomical Observatory of Japan, 2-21-1 Osawa, Mitaka, Tokyo, 181-8588, Japan}
\affiliation{Department of Physics, Faculty of Science and Engineering, Konan University, 8-9-1 Okamoto, Kobe, Hyogo 658-8501, Japan}

\author[0000-0002-6047-430X]{Yuichi Harikane}
\affiliation{Institute for Cosmic Ray Research, The University of Tokyo, 5-1-5 Kashiwa-no-Ha, Kashiwa, Chiba, 277-8582, Japan}

\author[0000-0003-4656-0241]{Miho N. Ishigaki}
\affiliation{National Astronomical Observatory of Japan, 2-21-1 Osawa, Mitaka, Tokyo, 181-8588, Japan}

\author[0000-0001-7730-8634]{Yuki Isobe}
\affiliation{Kavli Institute for Cosmology, University of Cambridge, Madingley Road, Cambridge, CB3 0HA, UK}
\affiliation{Cavendish Laboratory, University of Cambridge, 19 JJ Thomson Avenue, Cambridge, CB3 0HE, UK}
\affiliation{Waseda Research Institute for Science and Engineering, Faculty of Science and Engineering, Waseda University, 3-4-1, Okubo, Shinjuku, Tokyo 169-8555, Japan}

\author[0009-0000-1999-5472]{Minami Nakane}
\affiliation{Institute for Cosmic Ray Research, The University of Tokyo, 5-1-5 Kashiwa-no-Ha, Kashiwa, Chiba, 277-8582, Japan}
\affiliation{Department of Physics, Graduate School of Science, The University of Tokyo, 7-3-1 Hongo, Bunkyo, Tokyo 113-0033, Japan}

\author[0000-0003-4321-0975]{Moka Nishigaki}
\affiliation{Department of Astronomical Science, SOKENDAI (The Graduate University for Advanced Studies), \\ 2-21-1 Osawa, Mitaka, Tokyo, 181-8588, Japan}
\affiliation{National Astronomical Observatory of Japan, 2-21-1 Osawa, Mitaka, Tokyo, 181-8588, Japan}

\author[0000-0001-9553-0685]{Ken'ichi Nomoto}
\affiliation{Kavli Institute for the Physics and Mathematics of the Universe (WPI), The University of Tokyo, Kashiwa, Chiba 277-8583, Japan}

\author[0000-0001-9011-7605]{Yoshiaki Ono}
\affiliation{Institute for Cosmic Ray Research, The University of Tokyo, 5-1-5 Kashiwa-no-Ha, Kashiwa, Chiba, 277-8582, Japan}

\author[0000-0003-3228-7264]{Masato Onodera}
\affiliation{Department of Astronomical Science, SOKENDAI (The Graduate University for Advanced Studies), \\ 2-21-1 Osawa, Mitaka, Tokyo, 181-8588, Japan}
\affiliation{National Astronomical Observatory of Japan, 2-21-1 Osawa, Mitaka, Tokyo, 181-8588, Japan}
\affiliation{Subaru Telescope, National Astronomical Observatory of Japan, National Institutes of Natural Sciences (NINS),\\ 650 North A’ohoku Place, Hilo, HI 96720, USA}

\author[0000-0002-7043-6112]{Akihiro Suzuki}
\affiliation{Research Center for the Early Universe, Graduate School of Science, The University of Tokyo, \\7-3-1 Hongo, Bunkyo, Tokyo 113-0033, Japan}

\author[0000-0002-6705-6303]{Koh Takahashi}
\affiliation{National Astronomical Observatory of Japan, 2-21-1 Osawa, Mitaka, Tokyo, 181-8588, Japan}

\author[0009-0005-2897-002X]{Yui Takeda}
\affiliation{Department of Astronomical Science, SOKENDAI (The Graduate University for Advanced Studies), \\ 2-21-1 Osawa, Mitaka, Tokyo, 181-8588, Japan}
\affiliation{National Astronomical Observatory of Japan, 2-21-1 Osawa, Mitaka, Tokyo, 181-8588, Japan}

\author[0009-0006-6763-4245]{Hiroto Yanagisawa}
\affiliation{Institute for Cosmic Ray Research, The University of Tokyo, 5-1-5 Kashiwa-no-Ha, Kashiwa, Chiba, 277-8582, Japan}
\affiliation{Department of Physics, Graduate School of Science, The University of Tokyo, 7-3-1 Hongo, Bunkyo, Tokyo 113-0033, Japan}

\begin{abstract}
We present chemical abundance ratios of 8 nitrogen-rich ([N/O]$>0.3$) galaxies at $z\sim 6-12$ identified by the first 4 years of the JWST observations, and compare these ratios with chemical evolution models. We reanalyze the JWST/NIRSpec data of these galaxies in the self-consistent manner for line fluxes and upper limits including those previously unconstrained. We derive the abundance ratios and constraints of [N/O], [C/O], [Ne/O], [Ne/C], [Ar/O], [S/O] and [Fe/O], characterizing the nebulae in the galaxies with the electron temperatures and densities measured with {\sc[Oiii]}$\lambda4363$ and {\sc[Oii]}$\lambda\lambda3727, 3729$ lines, respectively. We develop the chemical evolution models for the three major scenarios, Wolf-Rayet stars, supermassive stars, and tidal disruption events (TDEs) with the AGB star contribution, integrating the ejecta of the stars and core-collapse supernovae (CCSNe) over the age with yields calculated by numerical simulations. 
We compare the models with the [N/O] measurements and stellar ages, and find that all of the scenarios reproduce [N/O] as high as those of our galaxies. 
However, the time-scales of the high [N/O] ratios are too short to explain our galaxies in any of the scenarios, suggestive of very frequent failed supernovae that do not increase oxygen against nitrogen. We find that the three scenarios are distinguished in the plane of [Ne/C] vs. [N/O] due to Ne production outside CNO cycle, and that the observed abundance ratios are explained by the Wolf-Rayet models better than supermassive-star and TDE models. We argue that abundance ratios of various elements and time scales are clues for understanding nitrogen-rich galaxies.

\end{abstract}


\section{Introduction}
Understanding the abundance ratios of early galaxies is crucial for deciphering the chemical evolution processes that occur during galaxy formation. 
However, before the James Webb Space Telescope (JWST) was launched, it was challenging to investigate the abundance ratios of galaxies at high redshift due to observational limitations. 
JWST/Near Infrared Spectrograph (NIRSpec) allows us to observe emission lines in a near-infrared wavelength range of $1–5~\mathrm{\mu m}$ from high redshift ($z\gtrsim 6$) galaxies, 
which corresponds to the rest-frame ultraviolet (UV) to optical range.

One notable example of high redshift galaxies is GN-z11 at $z=10.6$ \citep{2016ApJ...819..129O}.
\cite{Bunker_2023} identify strong nitrogen emission lines of GN-z11 with JWST/NIRSpec data.
\cite{2023MNRAS.523.3516C} claim that GN-z11 has an extremely high N/O larger than the solar abundance, far exceeding that of nearby galaxies with similar metallicities.
The high N/O ratios in high-$z$ galaxies like GN-z11 have also been reported in recent studies \red{\citep{2023ApJ...959..100I,2025ApJ...980..225T,2024MNRAS.529.3301T, Schaerer_2024,2024MNRAS.535..881J,2024ApJ...972..143C}.}
\red{Similar galaxies with high N/O abundance ratios have also been reported at lower redshifts \citep{2024ApJ...964L..12R,2024A&A...690A.269R,2025MNRAS.540.2991A,2025ApJ...981..136S,2024ApJ...962...24S}.}
The abundance ratios observed in high N/O galaxies resemble those found in stars within globular clusters (GCs).
These high N/O galaxies at high redshift may be progenitors of GCs \citep{2024ApJ...966...92S}.
Chemical evolution models developed by \cite{N13papaer}, including contributions from AGB stars, cannot reproduce the unusually high N/O ratios observed \citep{2024ApJ...962...50W}.
Since the nitrogen enrichment from AGB stars occurs on a longer timescale ($\gtrsim$ 70 Myr) compared to the oxygen enrichment from core-collapse supernovae (CCSNe), AGB stars cannot account for the elevated N/O ratios observed in these high-N/O galaxies at early epochs.
\cite{2023ApJ...959..100I} report that the N, O, and C ratios of high N/O galaxies are close to the CNO cycle equilibrium and suggest that these galaxies are enriched by the CNO cycle.
Three scenarios have been proposed as the origin of nitrogen related to the CNO cycle: Wolf-Rayet (WR) stars, supermassive stars (SMS), and tidal disruption events (TDE).
The contribution of nitrogen-rich stellar winds from WR stars has been extensively discussed in recent literature \citep[e.g.,][]{2024ApJ...962...50W, 2024ApJ...962L...6K, 2024PASJ...76.1122F}.
Recent studies have also debated the role of SMS ($\geq 10^3~M_\odot$) and TDE as the origin of the observed nitrogen enhancement \citep[e.g.,][]{2023ApJ...949L..16N, 2023A&A...673L...7C, 2025ApJ...994L..11N, 2026arXiv260104344E, 2023MNRAS.523.3516C}.
While the enrichment from AGB stars as the origin of high N/O galaxies has been ruled out, other scenarios involving WR stars, SMS, and TDE remain viable candidates. These scenarios remain indistinguishable due to the lack of elemental ratio indicators other than C/O.
\cite{2024ApJ...962...50W} develop chemical evolution models including WR stars and compare the abundance ratios of GN-z11.
The WR models reproduce the N/O ratio of GN-z11, but the N/O decreases in less than 1 Myr due to oxygen enrichment from CCSNe.
\cite{2024ApJ...962...50W} argue that the failed supernovae of massive stars are necessary to keep high N/O ratios.

In this study, we present the abundance ratios of high N/O galaxies at $z = 6-12$ observed with JWST/NIRSpec and discuss the abundance ratios of these galaxies with the chemical evolution models.
We constrain abundance ratios other than N/O and C/O through a homogeneous analysis of the sample and compare them with our chemical evolution model, which incorporates the failed supernova scenario suggested by \cite{2024ApJ...962...50W}.
Our high-$z$ galaxies and data reduction methods are described in Section \ref{sec:data_sample}.
Section \ref{sec:Analysis} shows our data analysis.
In Section \ref{sec:Nmodels}, we develop chemical evolution models of galaxies.
We present our results, and discuss the abundance ratios of our galaxies by comparing them with the chemical evolution models in Section \ref{sec:result}.
In Section \ref{sec:summary}, we summarize our study.

Throughout this paper, we assume a solar metallicity $Z_\odot$ as $12+\log({\rm O/H}) = 8.69$, and use the solar abundance ratios of $\log({\rm N/O})=-0.86$, $\log({\rm C/O})=-0.23$, $\log({\rm Ar/O})=-2.31$, $\log({\rm Ne/O})=-0.63$,  $\log({\rm S/O})=-1.57$, and $\log({\rm Fe/O})=-1.23$  \citep{Asplaud}.
Abundance ratios are defined by those normalized by the solar abundance ratios,
\begin{equation}
    \mathrm{[A/B]} = \log_{10} \left( \frac{N_A / N_{A,\odot}}{N_B / N_{B,\odot}} \right),
\end{equation}
where $N_A$ and $N_B$ are the numbers of the element A and B, respectively. The variables of $N_{A, \odot}$ and $N_{B,\odot}$ indicate the solar abundances.

\section{Data and Sample} \label{sec:data_sample}
This study needs galaxies with high N/O at $z = 6-12$ to investigate the origin of rich nitrogen in the early universe.
\red{We therefore select galaxies that have been identified as nitrogen-rich ([N/O]$>0.3$) based on rest-frame UV nitrogen emission lines. 
Specifically, we require detections of {\sc Niv]} $\lambda\lambda1483,1486$ and/or {\sc Niii]} $\lambda1750$, together with ancillary oxygen and hydrogen lines needed to derive or constrain N/O in a consistent abundance framework.}
We focus on bright objects ($M_{UV}<-19$) observed with the high- or medium-resolution gratings ($R \sim 1000 - 2700$) and PRISM ($R\sim100$) of the JWST/NIRSpec. 
These observations provide high signal-to-noise ratios, which are essential for detecting faint emission lines.
\red{We do not include galaxies whose nitrogen abundances are based only on rest-frame optical {\sc [Nii]} emission, such as object 10010 in \citet{2025ApJ...981..136S} and object 397 in \citet{2024ApJ...962...24S}, because {\sc [Nii]} traces a different ionization zone and requires a different ICF treatment from the UV {\sc Niv]}/{\sc Niii]} diagnostics adopted here.}

\begin{figure*}
    \centering
    \includegraphics[width=18cm]{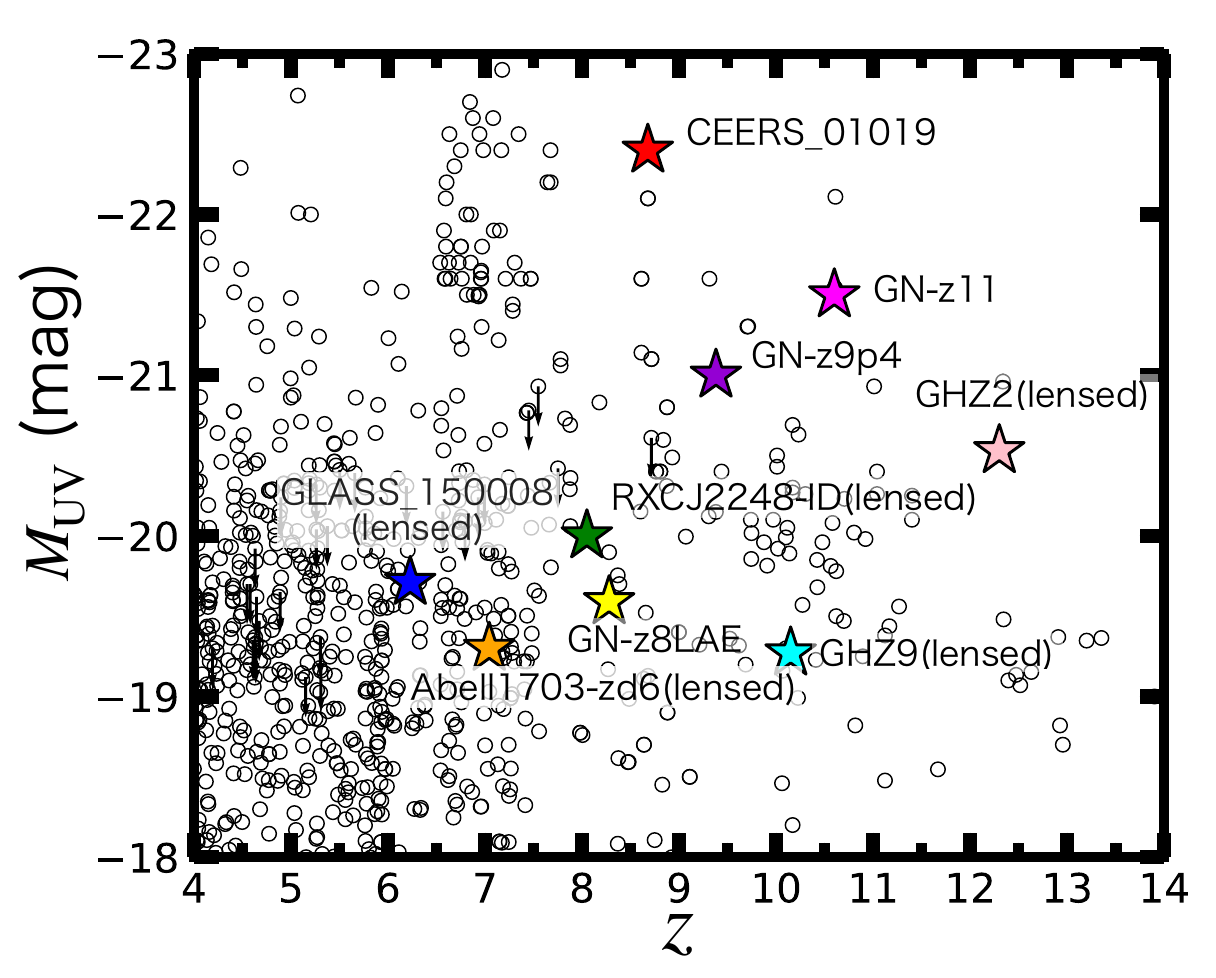}
    \caption{Redshift and UV magnitude of our high-$z$ galaxies. The star symbols represent our high-$z$ galaxies \citep{Bunker_2023, 2023ApJS..269...33N,2024MNRAS.529.3301T,2025ApJ...980..225T,  Schaerer_2024,2024ApJ...972..143C,2025ApJ...989...75N,2025ApJ...993..194N}. The open circles show galaxies that have been observed with JWST in the literature \citep{2023ApJS..269...33N, 2025ApJS..277....4D, 2024ApJ...960...56H,2025ApJ...980..138H,2025arXiv250821708R}.}
    \label{fig:MUV}
\end{figure*}

\begin{figure*}
    \centering
    \includegraphics[width=18cm]{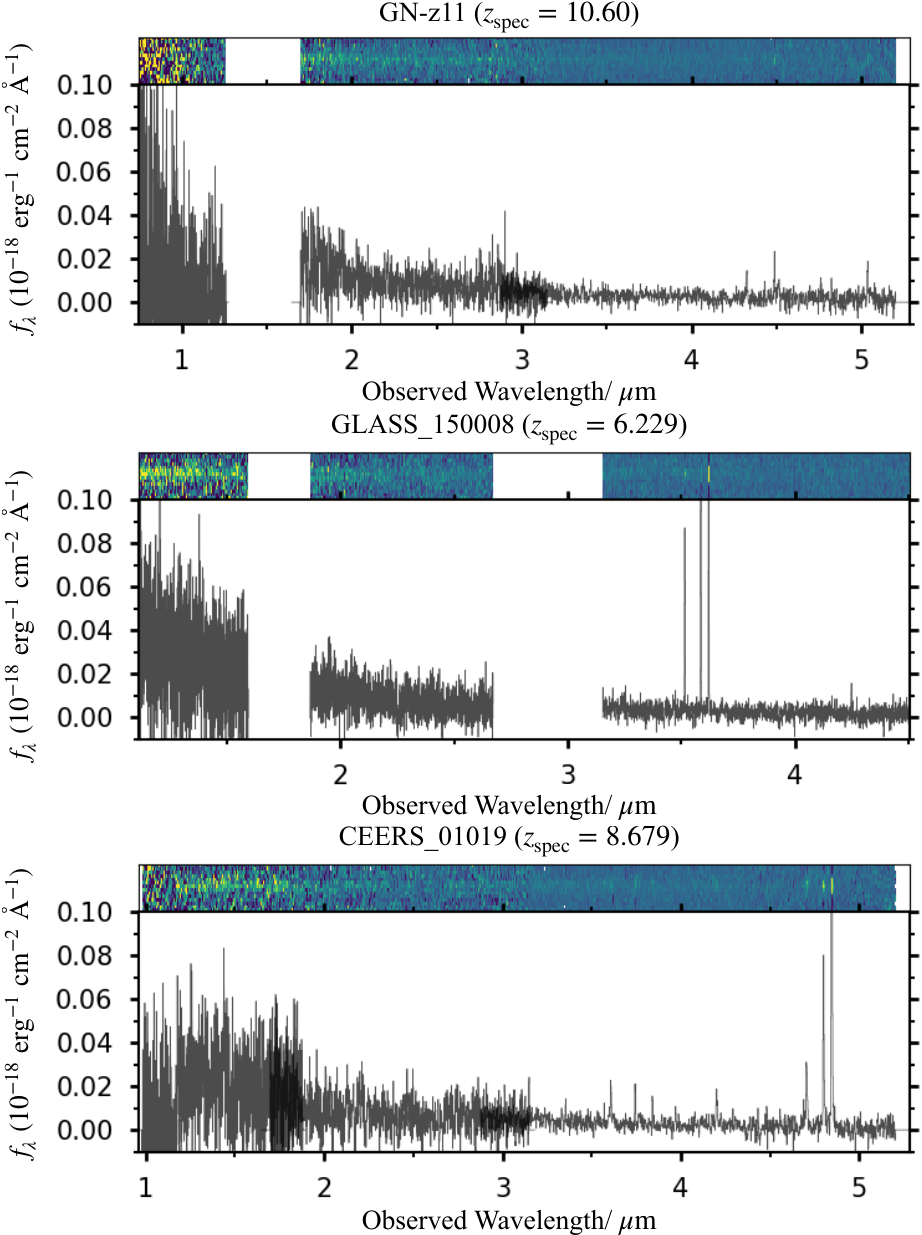}
    \caption{
    Grating spectra of GN-z11, GLASS\_150008, and CEERS\_01019 (from top to bottom). 
    Within each panel, the 2D spectrum is displayed at the top with the corresponding extracted 1D spectrum shown below.}
    \label{fig:obs_1}
\end{figure*}

\begin{figure*}
    \centering
    \includegraphics[width=18cm]{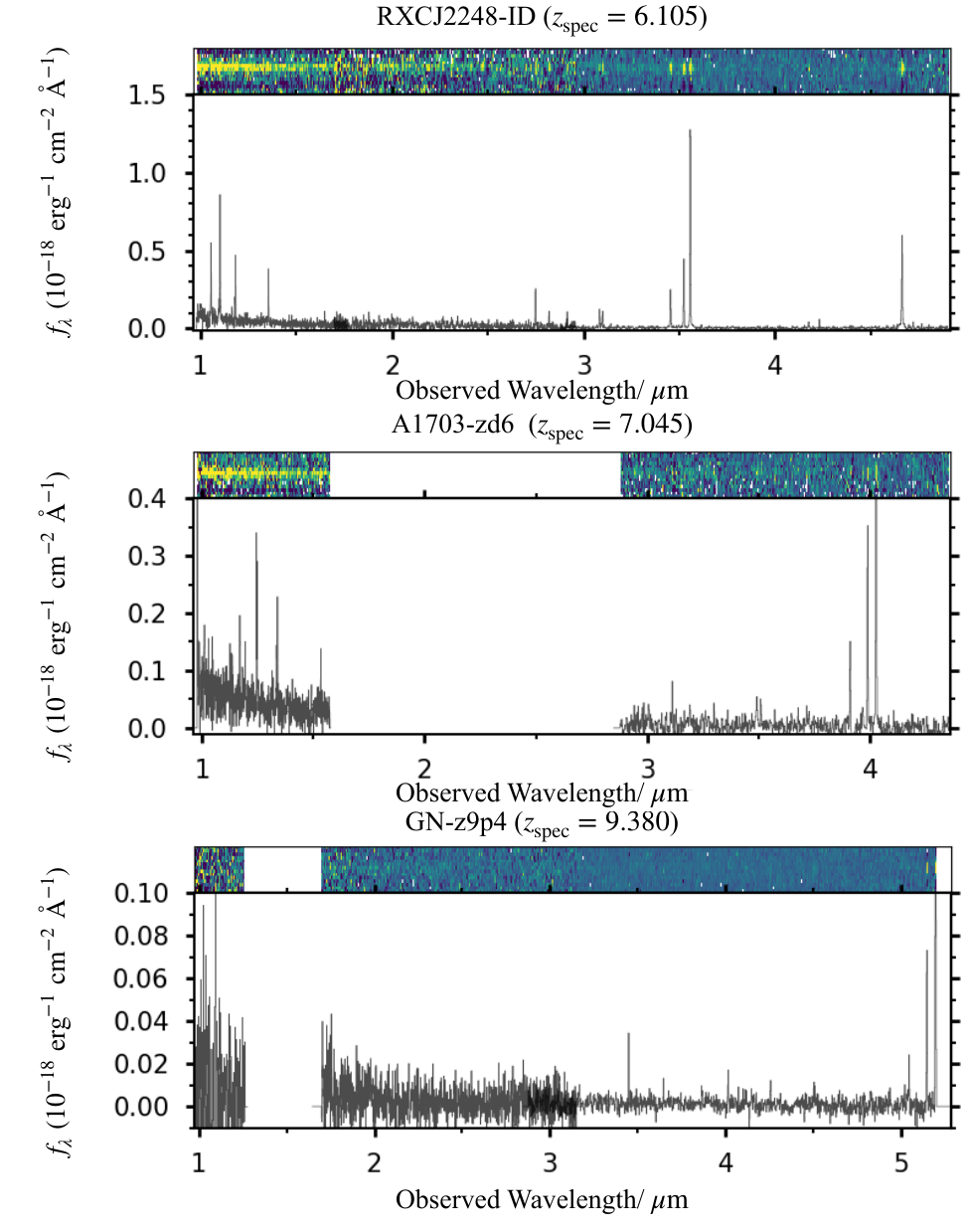}
    \caption{Same as Figure \ref{fig:obs_1}, but for RXCJ2248-ID(top) and A1703-zd6(middle), GN-z9p4(bottom).}
    \label{fig:obs_2}
\end{figure*}

\begin{figure*}
    \centering
    \includegraphics[width=18cm]{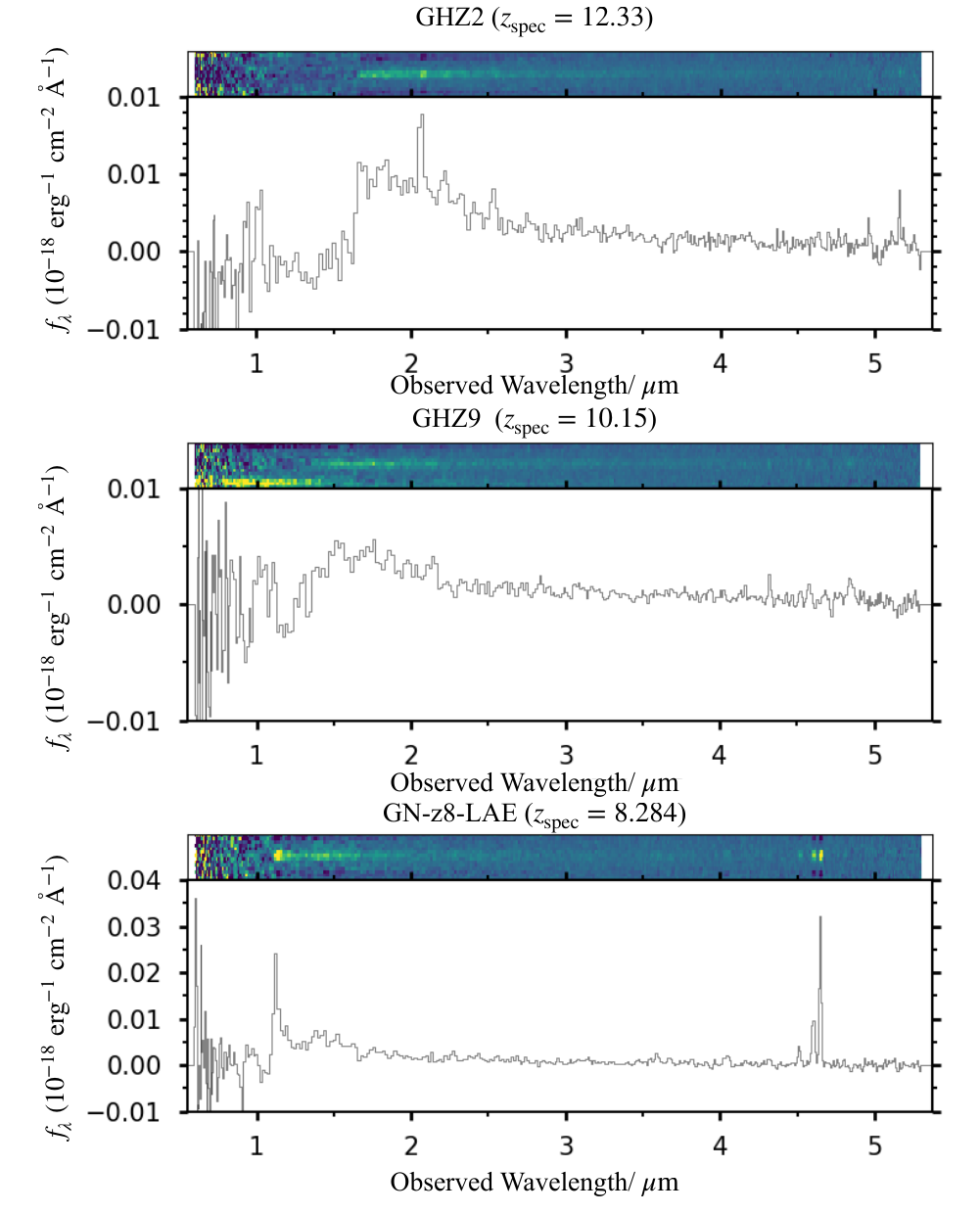}
    \caption{Prism spectra of GHZ2, GHZ9, and  GN-z8-LAE (from top to bottom). Within each panel, the 2D spectrum is displayed at the top with the corresponding extracted 1D spectrum shown below.}
    \label{fig:obs_3}
\end{figure*}

We analyze JWST/NIRSpec data of GN-z11 \citep{Bunker_2023}, GLASS\_150008, CEERS\_01019 \citep{2023ApJ...959..100I}, RXCJ2248-ID \citep{2024MNRAS.529.3301T}, A1703-zd6 \citep{2025ApJ...980..225T}, GN-z9p4 \citep{Schaerer_2024}, GHZ2 \citep{2024ApJ...972..143C}, GHZ9 \citep{2025ApJ...989...75N}, and GN-z8-LAE \citep{2025ApJ...993..194N}. 
While these sources are known for their characteristically high N/O ratios ([N/O] $> 0.3$), we perform a systematic re-evaluation of their N/O and other elemental abundance ratios to investigate their nitrogen origins.
We define our high-$z$ galaxies that consist of a total of the 9 galaxies, and summarize our high-$z$ galaxies in Table \ref{table_taget}.
Figure \ref{fig:MUV} shows the redshift and UV magnitudes of our high-$z$ galaxies and galaxies previously observed with JWST. 
Our high-$z$ galaxies consist of galaxies that are particularly bright or gravitationally lensed.
The reduced spectra of our galaxies are shown in Figures \ref{fig:obs_1} to \ref{fig:obs_3}.

\subsection{GN-z11} \label{sec:GN-z11}
GN-z11 in the GOODS-North field was observed with JWST/NIRSpec as part of the JWST Advanced Deep Extragalactic Survey (JADES; PID: 1181, PI: D. Eisenstein).
The data of GN-z11 were taken with the medium-resolution ($R\sim 1000$) gratings and filters of G140M/F100LP, G235M/F170LP, and G395M/F290LP and the low-resolution  ($R\sim 100$) PRISM/CLEAR mode.
The total exposure time for the medium-resolution and the low-resolution is 3100s and 6100s, respectively.

\subsection{GLASS\_150008}
The spectroscopic data of GLASS\_150008 were obtained in the GLASS program \citep{2022ApJ...935..110T}.
The GLASS program has targeted objects behind the galaxy cluster Abell 2744.
The GLASS data were obtained using $R\sim 2700$ high-resolution gratings/filters of G140H/F100LP, G235H/F170LP, and G395H/F290LP,  covering the wavelength ranges of 1.0–1.9, 1.7–3.1, and 2.9–5.1 $\mathrm{\mu m}$.
The total exposure time for each grating and filter pair is 4.9 hours.

\subsection{CEERS\_01019}
CEERS\_01019 was observed with JWST/NIRSpec in the CEERS \citep{2023ApJ...946L..13F} program.
The CEERS program was carried out in the EGS HST legacy field.
The CEERS data were obtained using the $R\sim 1000$ medium resolution gratings/filters of G140M/F100LP, G235M/F170LP, and G395M/F290LP, covering the wavelength ranges of 1.0–5.1 $\mathrm{\mu m}$.
The total exposure time for each prism, G140M, G235M, and G395M is 0.9 hours.

\subsection{RXCJ2248-ID}
RXCJ2248-ID is located in the RXCJ2248-4431 lensing field.
The spectroscopic data of RXCJ2248-ID were obtained with JWST/NIRSpec.
RXCJ2248-ID was observed in General Observers (GO) program, as part of Cycle 1 program ID 2478 (PI: D.Stark).
RXCJ2248-ID was taken with G140M/F100LP, G235M/F170LP, and G395M/F290LP gratings and filters.
The total exposure time for each grating and filter pair was 1.7 hours, 0.44 hours, and 0.44 hours, respectively.

\subsection{A1703-zd6}
JWST/NIRSpec spectroscopy of A1703-zd6 was conducted in Cycle 1 GO program ID 2478 (PI: D.Stark) targeting the Abell 1703 lensing field.
A1703-zd6 was observed with G140M/F100LP and G395M/F290LP gratings and filters with a total exposure time of 1.7 hours and 0.44 hours, respectively.

\subsection{GN-z9p4}
GN-z9p4 is a galaxy at $z=9.380$ in the GOODS-North field \citep{Schaerer_2024}.
GN-z9p4 was observed as part of the JADES under the same observational conditions as those of GN-z11 (Section \ref{sec:GN-z11}).

\subsection{GHZ2}
GHZ2 is located in the GLASS-JWST NIRCam field \citep{2022ApJ...935..110T}.
GHZ2 was observed with JWST/NIRSpec PRISM-CLEAR configuration ($R\sim 100$) in the Cycle 2 program GO-3073 (PI M. Castellano).
The total exposure time was 9.1 hours.

\subsection{GHZ9}
GHZ9 in the GLASS-JWST NIRCam field \citep{2022ApJ...935..110T} was identified by \cite{2023ApJ...948L..14C}.
GHZ9 was observed in the same program as GHZ2 (GO 3073).
The observations were conducted using the JWST/NIRSpec CLEAR-PRISM configuration with a total exposure time of 5.5 hours.

\subsection{GN-z8-LAE}
GN-z8-LAE is a strong Ly$\mathrm{\alpha}$ emitter at $z = 8.284$ in the GOODS-North field \citep{2025ApJ...993..194N}.
GN-z8-LAE was observed as part of the JADES under the same observational conditions as those of GN-z11 (Section \ref{sec:GN-z11}).

\subsection{Data Reduction}
We extract the data from the Mikulski Archive for Space Telescopes  (MAST) at the Space Telescope Science Institute and conduct level-2 and -3 calibrations for the spectroscopic data of our high-$z$ galaxies using the JWST Science Calibration Pipeline with the Calibration Reference Data System (CRDS) in the same manner as \cite{2023ApJS..269...33N}. 
See \cite{2023ApJS..269...33N} for the details of the data reduction.
\red{The JWST data presented in this paper were obtained from the MAST. The specific observations analyzed can be accessed via
\dataset[doi:10.17909/2egt-j990]{https://doi.org/10.17909/2egt-j990}.}

\begin{table*}[t]
    \begin{center}
    \tabletypesize{\scriptsize}
    \tablewidth{0pt} 
    \caption{Our Sample}
    \begin{tabular}{ccccc}\hline \hline
        ID & R.A. & Dec. & Redshift & Reference\\
           & hh:mm:ss & dd:mm:ss  &  & \\
     (1) & (2) & (3) & (4) & (5)\\ \hline
    GN-z11 & 12:36:25.46 & $+$62:14:31.4 & 10.60 & \cite{Bunker_2023}\\
    GLASS\_150008 & 00:14:24.6 & $-$30:25:9.2 & 6.229 & \cite{2023ApJ...959..100I} \\
    CEERS\_01019 & 14:20:8.5 & $+$52:53:26.4 & 8.679 & \cite{2023ApJ...959..100I} \\ 
    RXCJ2248-ID & 22:48:43 & $-$44:32:4.2 & 6.105 & \cite{2024MNRAS.529.3301T} \\ 
    A1703-zd6 & 13:15:3.8 & $+$51:49:18.3 & 7.045 & \cite{2025ApJ...980..225T}\\ 
    GN-z9p4 & 12:35:4.1 & $+$62:14:29.7 & 9.380 & \cite{Schaerer_2024}\\
    GHZ2 & 00:14:1.0 & $-$30:20:49.3 & 12.33 & \cite{2024ApJ...972..143C}\\
    GHZ9 & 00:13:54.9 & $-$30:20:43.9 & 10.15 & \cite{2025ApJ...989...75N}\\
    GN-z8-LAE & 12:36:47.5 & $+$62:15:25.06 & 8.284 & \cite{2025ApJ...993..194N}\\
    \hline
    \end{tabular}
    
\tablecomments{(1) ID. (2) Right ascension in J2000. (3) Declination in J2000. (4) Redshift. (5) Reference.}
\label{table_taget}
\end{center}
\end{table*}

\clearpage

\begin{figure*}
    \centering
    \includegraphics[width=18cm]{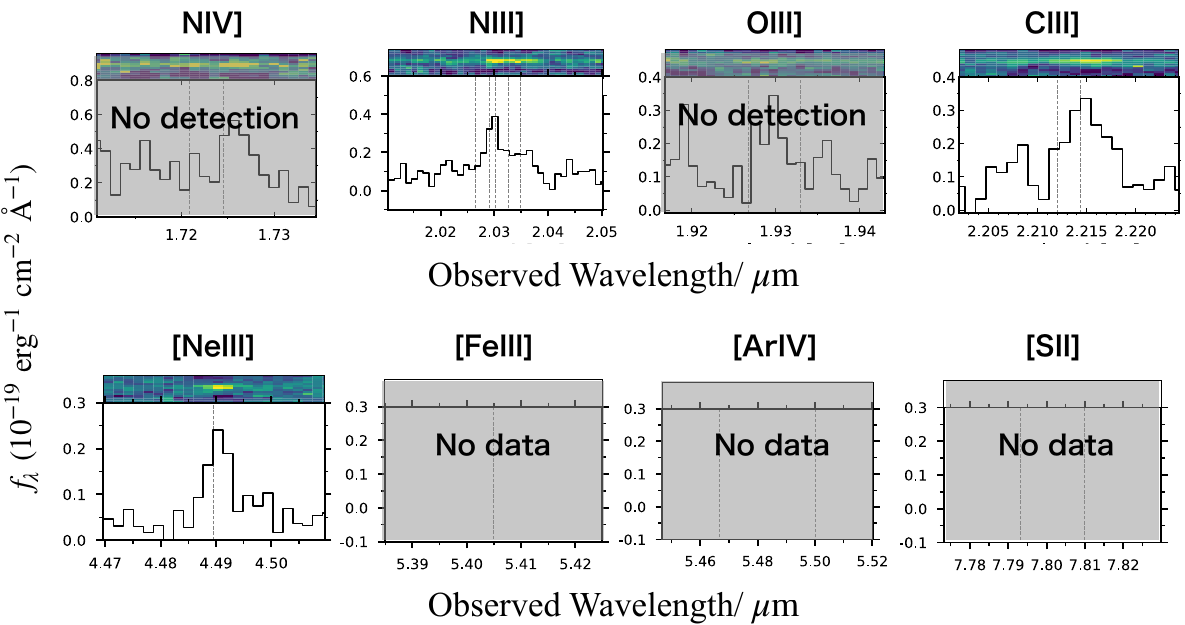}
    \caption{Emission-line detections of GN-z11. 
    These panels represent the enlarged view of the 1D and 2D spectrum around N{\sc iii}], N{\sc iv}], O{\sc iii}], C{\sc iii}], [Ne{\sc iii}], [Fe{\sc iii}], [Ar{\sc iv}], and [S{\sc ii}] (from top left to bottom right).
    The vertical dotted lines indicate the observed wavelength of each emission line.}
    \label{fig:emission}
\end{figure*}

\begin{figure*}
    \centering
    \includegraphics[width=18cm]{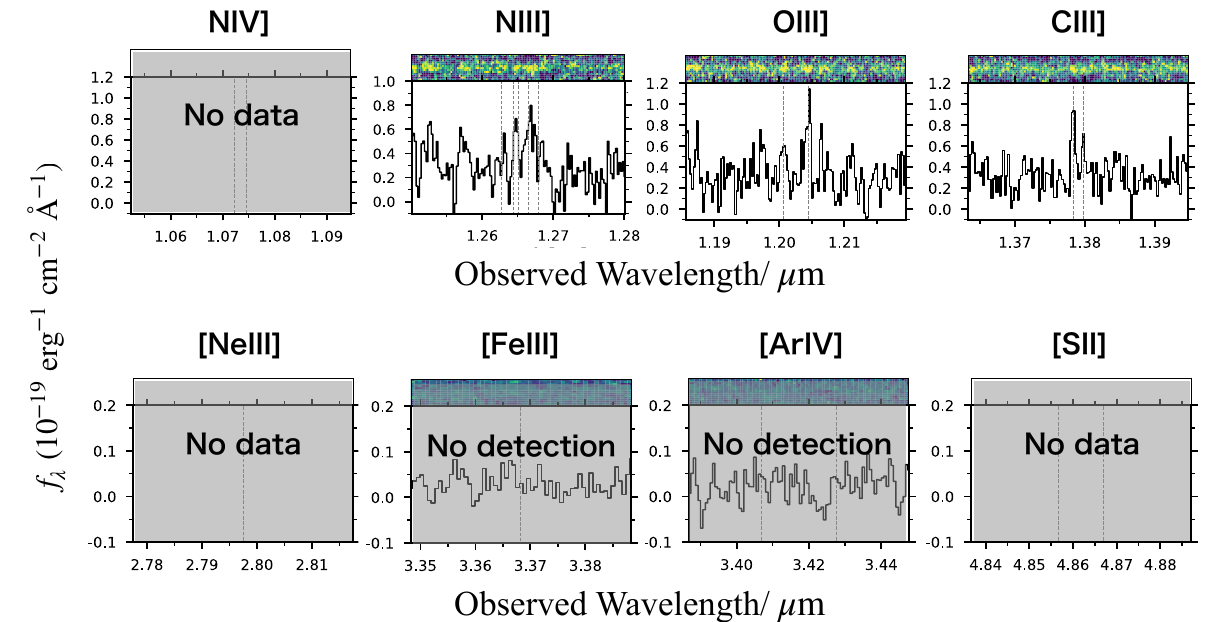}
    \caption{Same as Figure \ref{fig:emission}, but for GLASS\_150008.}
    \label{fig:emission_GLASS}
\end{figure*}

\begin{figure*}
    \centering
    \includegraphics[width=18cm]{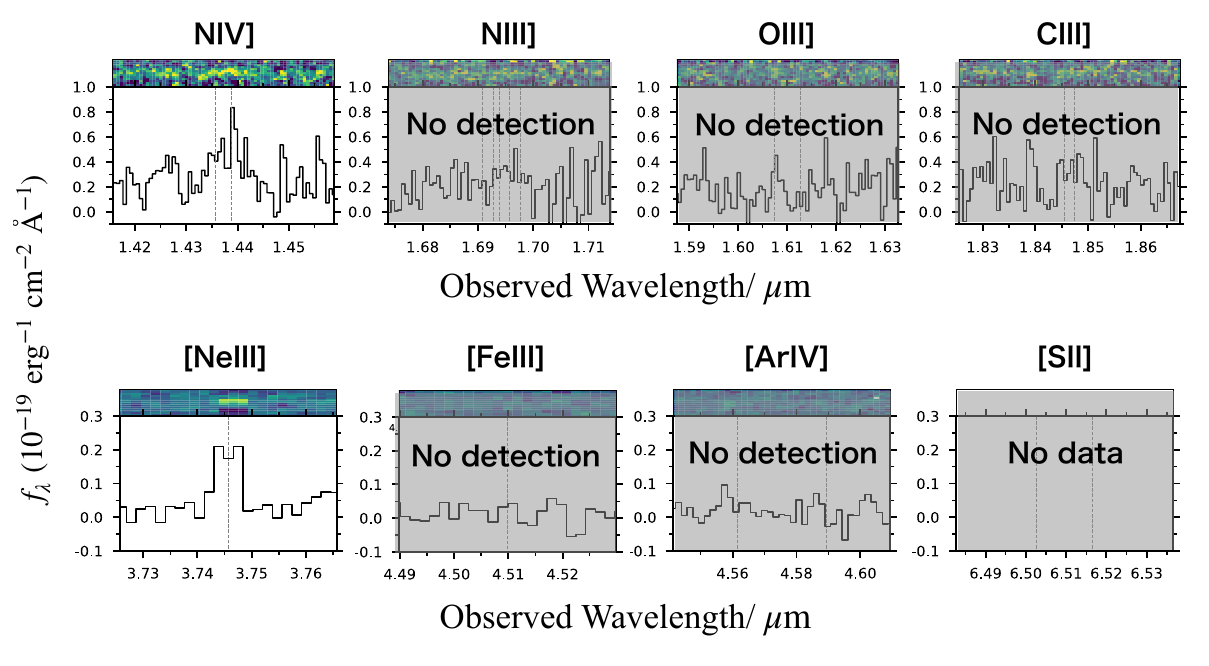}
    \caption{Same as Figure \ref{fig:emission}, but for CEERS\_01019.}
    \label{fig:emission_CEERS}
\end{figure*}

\begin{figure*}
    \centering
    \includegraphics[width=18cm]{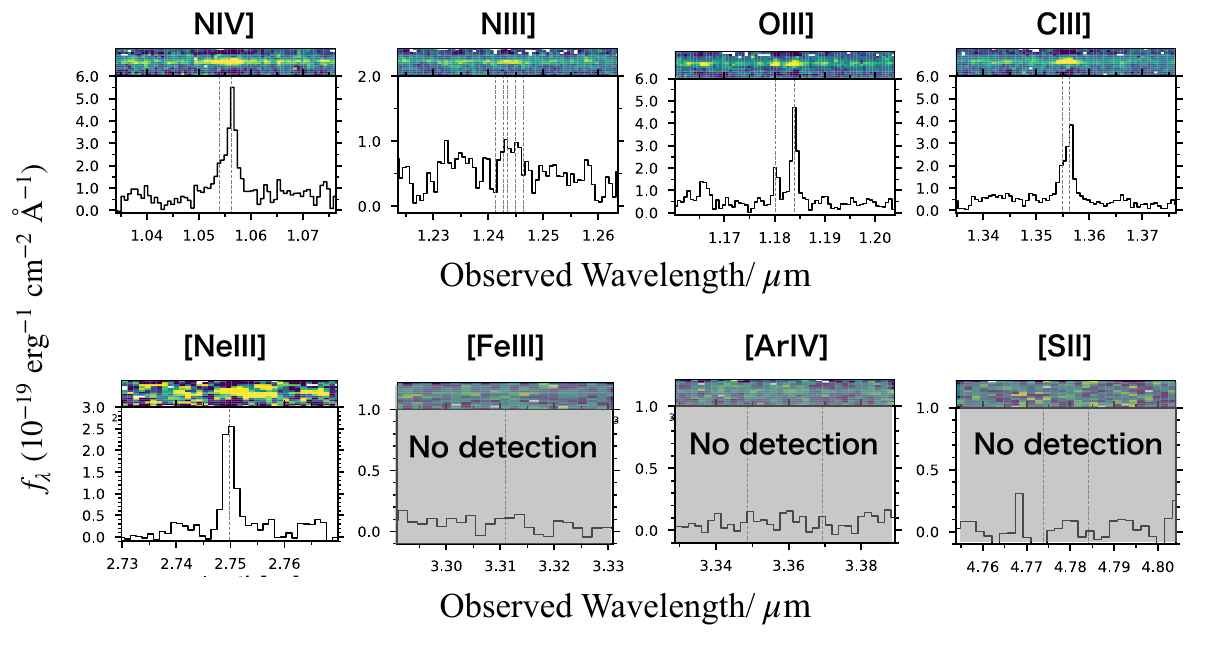}
    \caption{Same as Figure \ref{fig:emission}, but for RXCJ2248-ID.}
    \label{fig:emission_RX}
\end{figure*}

\begin{figure*}
    \centering
    \includegraphics[width=18cm]{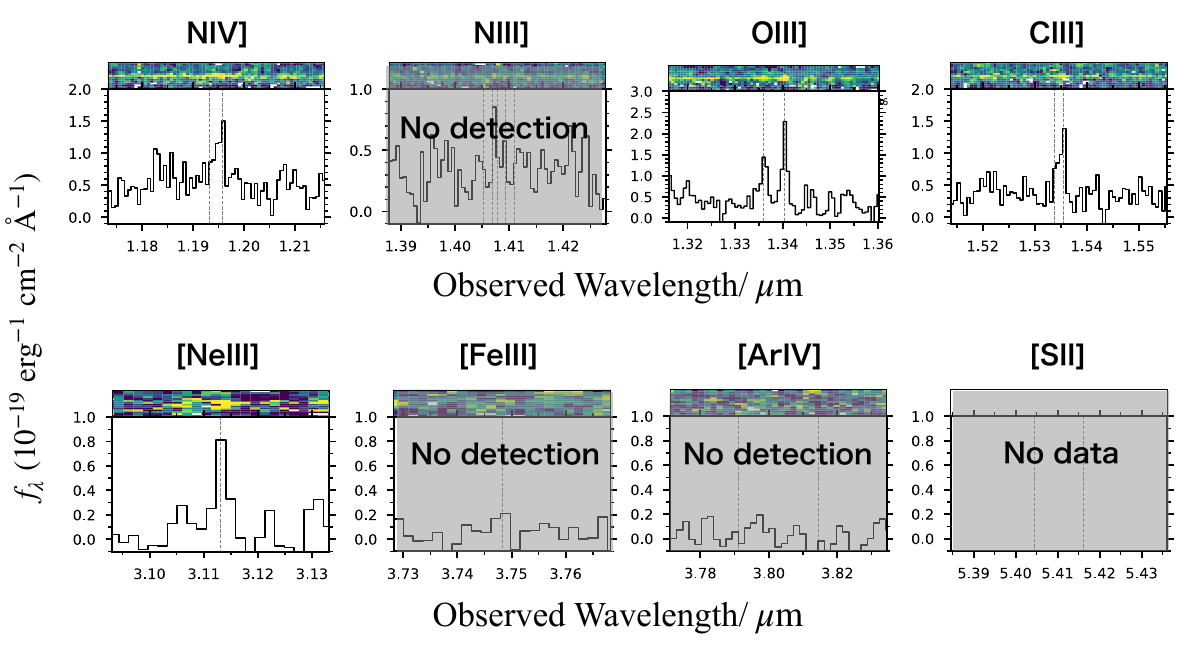}
    \caption{Same as Figure \ref{fig:emission}, but for A1703-zd6.}
    \label{fig:emission_Abell}
\end{figure*}

\begin{figure*}
    \centering
    \includegraphics[width=18cm]{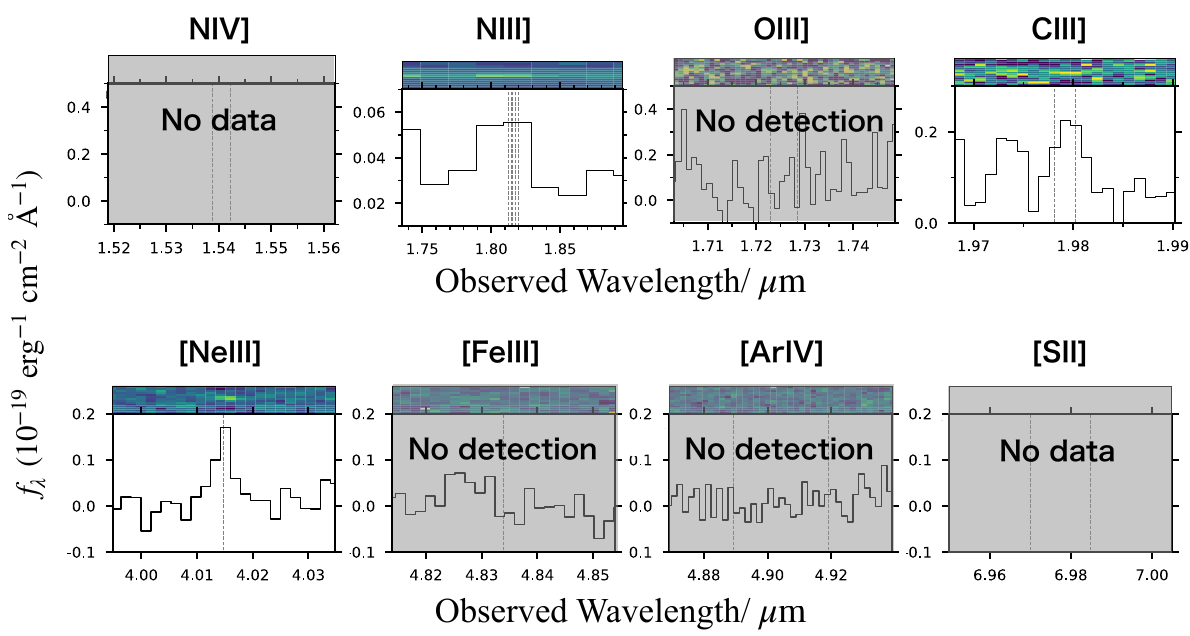}
    \caption{Same as Figure \ref{fig:emission}, but for GN-z9p4.
    The panel for N{\sc iii}] shows the spectrum obtained with the Prism configuration, as these lines are only detected in prism spectra; all other panels display Grating spectra.}
    \label{fig:emission_GNz9p4}
\end{figure*}

\begin{figure*}
    \centering
    \includegraphics[width=18cm]{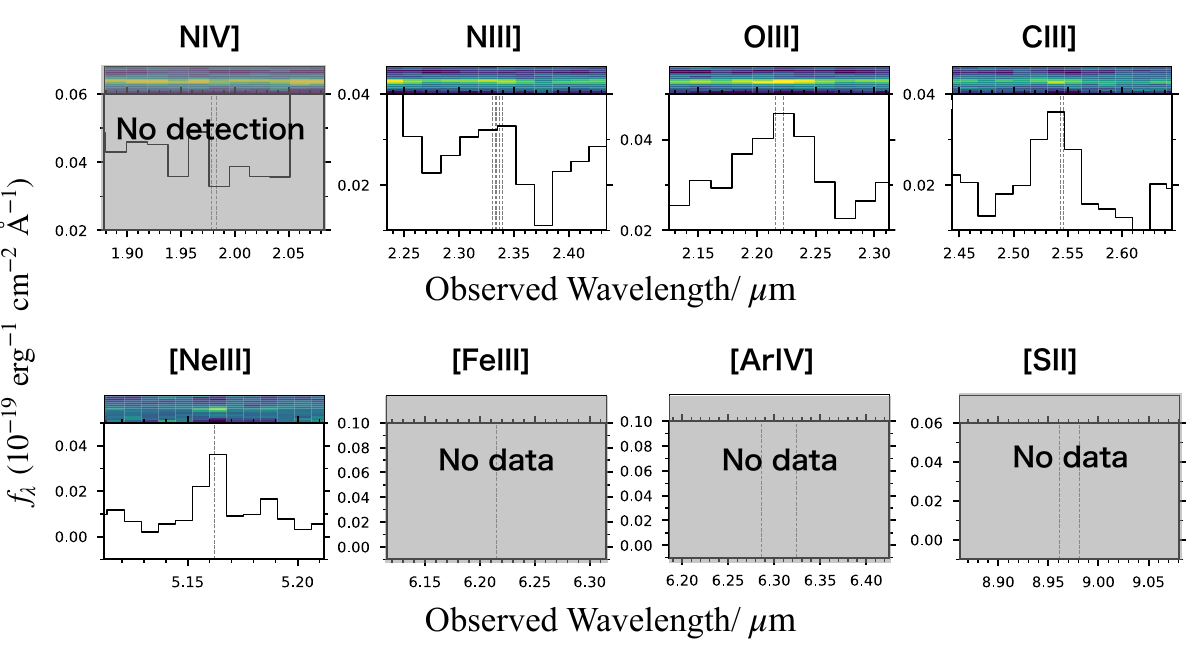}
    \caption{Same as Figure \ref{fig:emission}, but for GHZ2.}
    \label{fig:emission_GHZ2}
\end{figure*}

\begin{figure*}
    \centering
    \includegraphics[width=18cm]{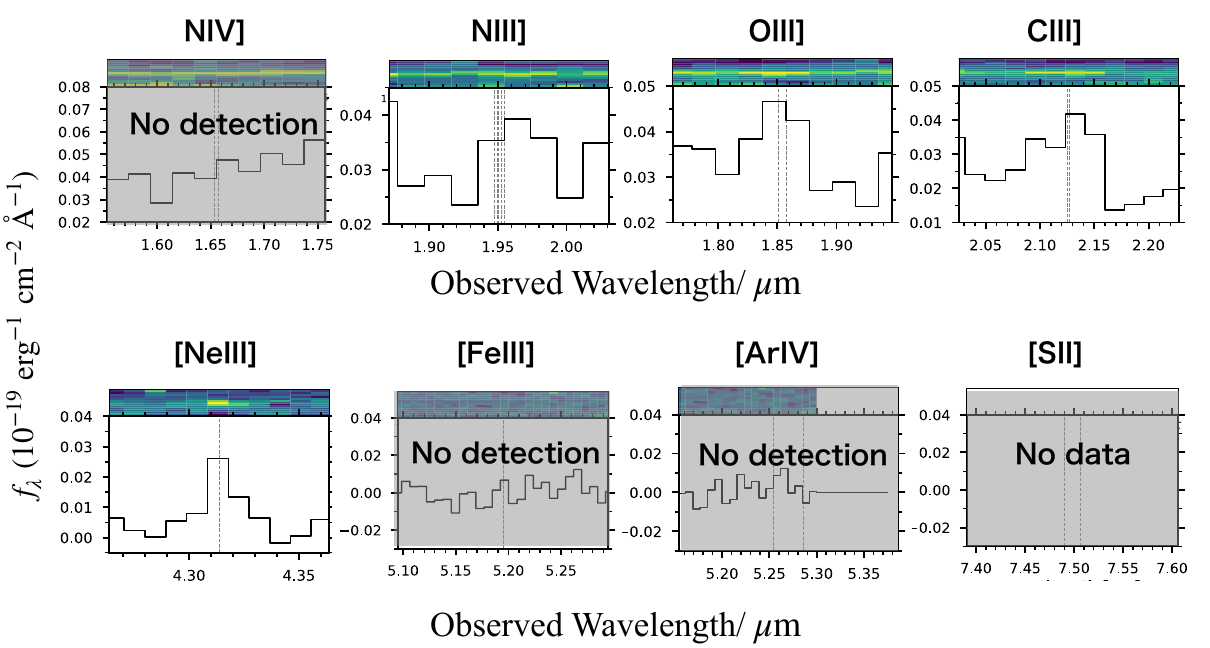}
    \caption{Same as Figure \ref{fig:emission}, but for GHZ9.}
    \label{fig:emission_GHZ9}
\end{figure*}

\begin{figure*}
    \centering
    \includegraphics[width=18cm]{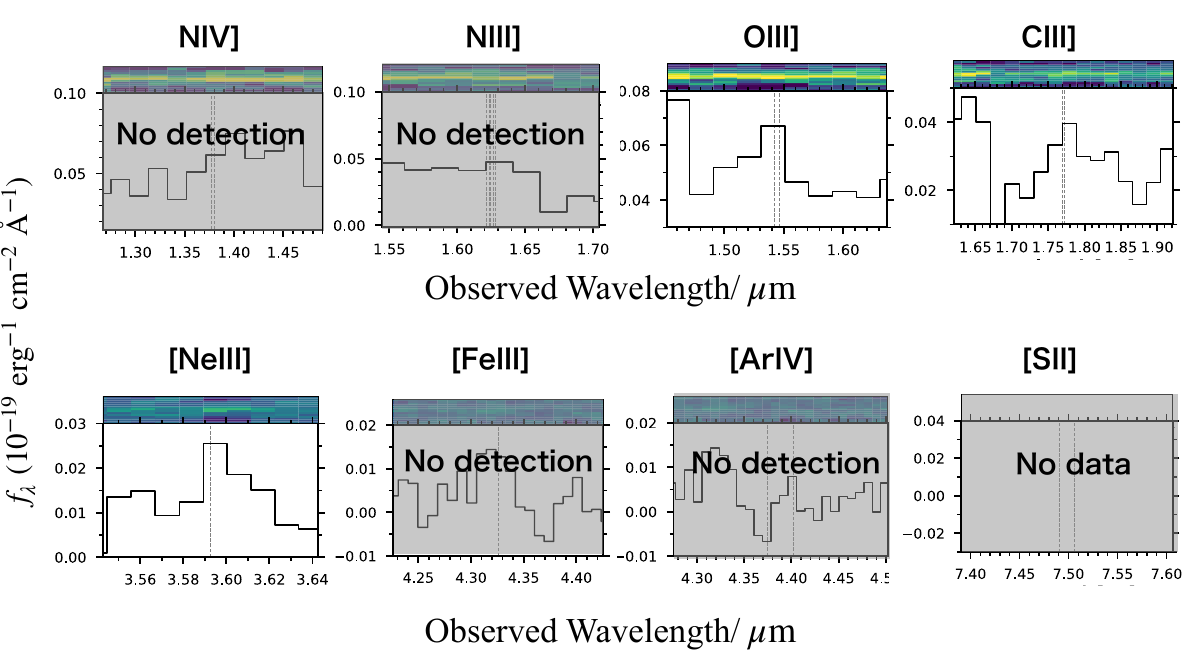}
    \caption{Same as Figure \ref{fig:emission}, but for GN-z8-LAE.}
    \label{fig:emission_LAE}
\end{figure*}

\section{Analysis} \label{sec:Analysis}
\subsection{Emission Line Measurements} \label{sec:emission_lines}
We measure the emission line flux by fitting a Gaussian profile convolved by line-spread functions provided by STScI \footnote{https://jwst-docs.stsci.edu/jwst-near-infrared-spectrograph/nirspecinstrumentation/nirspec-dispersers-and-filters} to account for
the instrumental broadening using the {\tt scipy.optimize} package \citep{2020NatMe..17..261V}.
We apply the $\chi^2$ minimization approach considering the error spectra.
We perform the emission-line fitting by fixing the redshift and velocity dispersion of the Gaussian profiles to those of the {\sc [Oiii]}$\lambda\lambda$4959,5007 lines.
Due to the redshifted {\sc [Oiii]}$\lambda\lambda$4959,5007 lines moving beyond the spectral coverage for $z > 10$ sources, we instead employed the [Ne{\sc iii]}$\lambda$3869 line for our velocity dispersion analysis. 
This line is selected as the primary kinematic tracer because it consistently provides the highest signal-to-noise ratio among the detected lines at these redshifts.
All wavelength measurements are based on the vacuum wavelength scale.
The error for each flux measurement is calculated by taking the sum of squares of the noise levels of spectral bins within a wavelength range of $\pm$ FWHM centered on the Gaussian peak.
For emission lines that are not detected, we calculate the 3$\sigma$ upper limits of fluxes.
Figures \ref{fig:emission}--\ref{fig:emission_LAE} present the spectra around the main emission lines for our high-$z$ galaxies. 
In Figure \ref{fig:emission_LAE}, the nitrogen emission lines of GN-z8-LAE, previously reported by \cite{2025ApJ...993..194N}, are not significantly detected. Therefore, we exclude GN-z8-LAE from our analysis to ensure the robustness of our results.

To perform reddening-corrections for the observed fluxes, we estimate dust extinction from the Balmer decrements under the assumptions of the case B \citep{1971MNRAS.153..471B} recombination and the dust attenuation curve given by \cite{2000ApJ...533..682C}.
We estimate intrinsic Balmer decrement values using PyNeb \citep{2015A&A...573A..42L}.
Because the Balmer decrement values depend on electron temperature $T_\mathrm{e}$ and electron density $n_\mathrm{e}$, 
we derive the color excesses $E(B-V)$ that consistently explain $T_\mathrm{e}$ and $n_\mathrm{e}$ (see Section \ref{sec:abundance} for the procedures of $T_\mathrm{e}$ and $n_\mathrm{e}$ calculations).
We utilize 6 Balmer line ratios within the observed wavelength range, H$\mathrm{\beta}$/H$\mathrm{\alpha}$, H$\mathrm{\gamma}$/H$\mathrm{\alpha}$, H$\mathrm{\delta}$/H$\mathrm{\alpha}$, H$\mathrm{\gamma}$/H$\mathrm{\beta}$, H$\mathrm{\delta}$/H$\mathrm{\beta}$, and H$\mathrm{\gamma}$/H$\mathrm{\delta}$ to estimate the emissivity of the Balmer lines.
We compare the Balmer line ratios of the observational measurements with those of theoretical predictions obtained with PyNeb to determine $E(B-V)$.
With the $E(B-V)$ values thus obtained,
we determine the best-estimate $E(B-V)$ by the $\chi^2$ minimization for each  Balmer decrement value.
We also estimate $\pm 68\ \%$ confidence intervals of $E(B - V)$ based on $\chi^2$.
With the $E(B - V )$ values and the attenuation curve \citep{2000ApJ...533..682C},
we correct all of the observed emission line fluxes for dust extinction. 
\red{For objects for which $E(B-V)$ is estimated from the Balmer decrements, the uncertainties of the intrinsic line fluxes include both the measurement uncertainties of the observed fluxes and the uncertainty associated with the dust correction, propagated from the uncertainty in $E(B-V)$.}

RXCJ2248-ID has been reported to show anomalies in the Balmer decrements \citep{2024ApJ...974..180Y, 2024MNRAS.tmp.1525C}.
Our analyzed spectra show similar anomalies in the Balmer decrements. Because it is difficult to reliably constrain the effects of dust extinction,  we assume $E(B-V) = 0$ for RXCJ2248-ID.

\subsection{Elemental Abundance Ratios} \label{sec:abundance}
We derive the oxygen abundances with the direct method for the galaxies in our sample where both {\sc [Oiii]} and {\sc [Oii]} emission lines are detected.
We use the PyNeb package {\tt getTemDen} to derive simultaneously  $T_\mathrm{e}(\mathrm{OIII})$ from emission-line ratios of {\sc [Oiii]}$\lambda$4363.
\red{For galaxies without a significant {\sc [Oiii]} $\lambda4363$ detection, we derive $T_{\rm e}({\rm O^{++}})$ using PyNeb from the UV-to optical ratio
\begin{equation}
R_{\rm O3,UV} =
\frac{
I({\sc Oiii]}\lambda1661)+I({\sc Oiii]}\lambda1666)
}{
I({\sc [Oiii]}\lambda4959)+I({\sc [Oiii]}\lambda5007)
}.
\label{eq:o3uv_ratio}
\end{equation}
}
We also derive $n_{\rm e}$ using the {\sc [Oii]} $\lambda\lambda3727,3729$
doublet.
For galaxies without a detection of the {\sc [Oii]} doublet, we assume $n_{\rm e}=1000~{\rm cm^{-3}}$. 
This value is consistent with the typical density found for galaxies at $z\sim7$--9 by \citet{2023ApJ...956..139I}.
\red{
For RXCJ2248-ID, we derive $n_{\rm e}$ from the {\sc Niv]} $\lambda\lambda1483,1486$ doublet, which has been shown to indicate a high electron density in this galaxy \citep{2025ApJ...980..225T}.
}
\red{This fiducial treatment applies the {\sc [Oii]}-based density to the ionic-abundance calculation, which is a simplifying assumption when low- and high-ionization lines are combined.
To assess the impact of a possible density difference, we perform a two-zone density-sensitivity calculation.
We keep the low-ionization species O$^+$, S$^+$, and Fe$^{2+}$ fixed at the fiducial density, and recompute the high-ionization species O$^{2+}$, N$^{2+}$, N$^{3+}$, C$^{2+}$, Ne$^{2+}$, and Ar$^{3+}$ over $n_{\rm e,high}=10^2$--$10^5~{\rm cm^{-3}}$.
The resulting abundance ratios show only modest changes over this density range, and the N/O enhancement remains in all cases.
Thus, the high N/O ratios are not driven by adopting the {\sc [Oii]}-based density for the high-ionization lines.}

We use the PyNeb package {\tt getIonAbundance} to obtain ionic abundance ratios.
The ionic abundance ratios O$^{2+}$/H$^+$ and O$^+$/H$^+$ are derived from
{\sc [Oiii]} $\lambda\lambda4959,5007$/H$\beta$ and/or
{\sc Oiii]} $\lambda\lambda1661,1666$/H$\beta$, and
{\sc [Oii]} $\lambda\lambda3727,3729$/H$\beta$, respectively.
For these calculations, we use $T_{\rm e}({\rm O^{++}})$ and $T_{\rm e}({\rm O^+})$, respectively.
The electron temperature of singly ionized oxygen is estimated using the empirical relation
\begin{equation}
T_{\rm e}({\rm O^+}) =
0.7 \times T_{\rm e}({\rm O^{++}}) + 3000
\end{equation}
\citep{1992AJ....103.1330G}.
By adding O$^{2+}$/H$^+$ and O$^+$/H$^+$, we obtain the total oxygen abundance O/H.
We express the oxygen abundance as $12+\log({\rm O/H})$.

For galaxies without {\sc [Oii]} detections, we derive oxygen abundances using the {\sc [Oiii]} $\lambda5007$/H$\beta$ ($R_3$) metallicity indicator.
We adopt the average $R_3$--metallicity relation presented by \citet{2022ApJS..262....3N}.
For GHZ2, the oxygen abundance is not determined because its Balmer lines fall outside the observed wavelength range.

We derive ion abundance ratios of N$^{3+}$/H$^+$,  N$^{2+}$/H$^+$, C$^{2+}$/H$^+$, Ne$^{2+}$/H$^+$, Ar$^{3+}$/H$^+$, S$^{+}$/H$^+$, and Fe$^{2+}$/H$^+$ that are estimated from the fluxes of 
{\sc Niv]}$\lambda\lambda$1483,1486, N{\sc iii}]$\lambda$1750 \footnote{In this paper, the N{\sc iii}] quintet centered at 1747, 1749, 1750, 1752, and 1754 $\textrm{\AA}$ is collectively referred to as N{\sc iii}]$\lambda$1750.}, {\sc Ciii]}$\lambda\lambda$1907,1909, [Ne{\sc iii]}$\lambda$3869, [Ar{\sc iv}]$\lambda$4711,4740, {\sc [Sii]}$\lambda\lambda$6716,6731, and [Fe{\sc iii]}$\lambda$4658, respectively, with the electron temperatures.
We adopt $T_{\mathrm{e}}(\text{O \sc iii})$ for $\text{N}^{2+}$, $\text{N}^{3+}$, $\text{C}^{2+}$, $\text{Ne}^{2+}$, and $\text{Ar}^{3+}$, because their ionization potentials ($41.0 - 77.5$ eV) are comparable to or higher than that of $\text{O}^{2+}$ (35.1 eV). 
For $\text{S}^{+}$ and $\text{Fe}^{2+}$, we apply $T_{\mathrm{e}}(\text{O \sc ii})$ because their ionization potentials ($16.2 - 23.3$ eV) are close to that of $\text{O}^{+}$ (13.6 eV), following the methodology of \cite{2021ApJ...922..170B}.

Due to the limited number of detected emission lines for GHZ2, we derived its ionic abundances using the following procedure \citep{2025arXiv250611846N}. 
Utilizing the latest atomic data implemented in PyNeb, we established analytical relations between the emission-line ratios and ion abundances.
The derived relations are expressed as follows:
\begin{equation}
\begin{split}
\log \left(\mathrm{N}^{2+} / \mathrm{O}^{2+}\right) &= \log \left(\frac{\mathrm{N} \text{\sc iii}] \lambda 1750}{\mathrm{O} \text{\sc iii}] \lambda \lambda 1661, 1666}\right) - 0.344 \\
&\quad  - \frac{0.252}{t_{\mathrm{e}}} - 0.034 t_{\mathrm{e}} + 0.118 \log \left(t_{\mathrm{e}}\right),
\end{split}
\end{equation}

\begin{equation}
\begin{split}
\log \left(\mathrm{C}^{2+} / \mathrm{O}^{2+}\right) &= \log \left(\frac{\mathrm{C} \text{\sc iii}] \lambda\lambda 1907,1909}{\mathrm{O} \text{\sc iii}] \lambda \lambda 1661, 1666}\right) - 0.781 \\
&\quad - \frac{0.527}{t_{\mathrm{e}}}   - 0.013 t_{\mathrm{e}} + 0.070 \log \left(t_{\mathrm{e}}\right),
\end{split}
\end{equation}

\begin{equation}
\begin{split}
\log \left(\mathrm{Ne}^{2+} / \mathrm{O}^{2+}\right) &= \log \left(\frac{[\mathrm{Ne} \text{\sc iii}] \lambda 3869}{\mathrm{O} \text{\sc iii}] \lambda \lambda 1661, 1666}\right) + 0.525 \\
&\quad - \frac{2.175}{t_{\mathrm{e}}}   - 0.034 t_{\mathrm{e}} + 0.118 \log \left(t_{\mathrm{e}}\right),
\end{split}
\end{equation}
where $t_e$ is the electron temperature of [{\sc Oiii]} in units of $10^4~\mathrm{K}$. 
These relations are valid over a wide temperature range ($0.4 < t_e < 5.0$) and assume an electron density of $n_e = 300~\mathrm{cm^{-3}}$. 
It should be noted that these relations remain nearly invariant for electron densities up to approximately $10^5~\mathrm{cm^{-3}}$.
Since the electron temperature for GHZ2 cannot be directly determined from the current data, we estimate the range of possible ion abundances by assuming $T_e$ values between $10,000$ and $25,000~\mathrm{K}$. 
This temperature range is adopted based on the results for high-redshift galaxies reported by \cite{2023ApJS..269...33N}.

We use the ionization correction factors (ICFs) to calculate total gas-phase elemental abundances from the ion abundances.
Recent studies suggest the presence of an AGN in GN-z11, CEERS\_01019, and GHZ9 \citep{2024Natur.627...59M, 2023ApJ...959..100I, 2025ApJ...989...75N}. To account for the impact of an AGN on the derived abundance ratios, we calculate the abundance ratios using two distinct ICFs: one assuming stellar radiation and the other assuming AGN radiation.
ICFs for both stellar and AGN radiation are derived via photoionization modeling with {\tt CLOUDY} \citep{2013hbic.book.....F}.
For the stellar-driven ICFs, we assume a young star-forming galaxy with a binary stellar population from BPASS \citep{2018MNRAS.479...75S}, following the parameters adopted by \cite{2023ApJ...959..100I}.
We employ an instantaneous star formation history with a stellar age of 10 Myr, a Salpeter initial mass function \citep[IMF; ][]{1955ApJ...121..161S}, and a hydrogen density of $n_\mathrm{H} = 300~\mathrm{cm^{-3}}$.
To derive the AGN ICFs, we utilize the {\tt CLOUDY} photoionization models presented in \cite{2018A&A...612A..94N}. 
We assume a fiducial parameter set: a power-law index $\alpha=-1.6$ and a Big Bump temperature $T_{\rm bb}=10^5$ K, which characterizes thermal emission from the accretion disc. 
The gas-phase metallicity is fixed to $0.2~Z_\odot$, a typical value observed in the three sources exhibiting AGN signatures.

\red{We compare our ICFs with the commonly used prescriptions of \citet{2006A&A...448..955I} and \citet{2021MNRAS.505.2361A} where applicable. 
A direct comparison for N/O is not possible because these literature nitrogen ICFs are based on the low-ionization optical ion N$^+$, whereas our nitrogen abundances are derived from the high-ionization UV lines {\sc Niii]} and {\sc Niv]}. 
For elements with comparable ionic prescriptions, we find that Ne/O is nearly insensitive to the choice of ICF, with differences smaller than 0.07 dex. 
The differences for S/O and Fe/O can reach a few tenths of a dex. 
For Ar/O, we cannot make a direct comparison with the \citet{2006A&A...448..955I} and \citet{2021MNRAS.505.2361A} prescriptions because our spectra do not provide  [Ar{\sc iii}] measurements. Those prescriptions require, or are calibrated for, argon ionic combinations including Ar$^{2+}$, whereas our Ar/O estimates are based only on Ar$^{3+}$ from [Ar{\sc iv}].
These ICF comparisons do not affect our main conclusions.}

\begin{equation}
\frac{\mathrm{N}}{\mathrm{H}}=\frac{\mathrm{N}^{3+}}{\mathrm{H}^{+}} \times \operatorname{ICF}\left(\mathrm{N}^{3+}\right),
\label{eq:N_H}
\end{equation}

\begin{equation}
\frac{\mathrm{N}}{\mathrm{H}}=\frac{\mathrm{N}^{2+}}{\mathrm{H}^{+}} \times \operatorname{ICF}\left(\mathrm{N}^{2+}\right),
\label{eq:N+_H}
\end{equation}

\begin{equation}
\frac{\mathrm{N}}{\mathrm{H}}=\frac{\mathrm{N}^{3+}+\mathrm{N}^{2+}}{\mathrm{H}^{+}} \times \operatorname{ICF}\left(\mathrm{N}^{3+}+\mathrm{N}^{2+}\right),
\label{eq:N++_H}
\end{equation}

\begin{equation}
\frac{\mathrm{C}}{\mathrm{H}}=\frac{\mathrm{C}^{2+}}{\mathrm{H}^{+}} \times \operatorname{ICF}\left(\mathrm{C}^{2+}\right),
\label{eq:C_H}
\end{equation}

\begin{equation}
\frac{\mathrm{Ne}}{\mathrm{H}}=\frac{\mathrm{Ne}^{2+}}{\mathrm{H}^{+}} \times \operatorname{ICF}\left(\mathrm{Ne}^{2+}\right),
\label{eq:Ne_H}
\end{equation}

\begin{equation}
\frac{\mathrm{Ar}}{\mathrm{H}}=\frac{\mathrm{Ar}^{3+}}{\mathrm{H}^{+}} \times \operatorname{ICF}\left( \mathrm{Ar}^{3+}\right),
\label{eq:Ar_H}
\end{equation}

\begin{equation}
\frac{\mathrm{S}}{\mathrm{H}}=\frac{\mathrm{S}^{+}}{\mathrm{H}^{+}} \times \operatorname{ICF}\left(\mathrm{S}^{+}\right),
\label{eq:S_H}
\end{equation}

\begin{equation}
\frac{\mathrm{Fe}}{\mathrm{H}}=\frac{\mathrm{Fe}^{2+}}{\mathrm{H}^{+}} \times \operatorname{ICF}\left(\mathrm{Fe}^{2+}\right).
\label{eq:Fe_H}
\end{equation}

Dividing the values of Eqs. (\ref{eq:N_H}) - (\ref{eq:Fe_H}) by O/H, we obtain N/O, C/O, Ne/O, Ar/O, S/O, and Fe/O.
For galaxies where both N{\sc iv}] and N{\sc iii}] lines are detected, we calculate the abundance ratios using Equation (\ref{eq:N++_H}).
\red{Uncertainties in the elemental abundance ratios are derived through a Monte Carlo approach. We generate 1,000 sets of intrinsic flux values by adding random Gaussian noise proportional to the $1\sigma$ intrinsic-flux errors, which include the contribution from the dust-correction uncertainty as described in Section\ref{sec:emission_lines}. The median and $\pm68\%$ confidence intervals are adopted as the representative values and their errors.}
For our galaxies, where the emission lines of N, O, C, Ne, Ar, S, and Fe are not detected, we use their 3$\sigma$ flux upper limits to place corresponding upper (lower) limits on the abundance ratios.

\red{Table~\ref{table_Tene} summarizes the adopted $T_{\rm e}({\rm O^{++}})$ values and the diagnostics used to estimate O/H for each galaxy.
This table clarifies which galaxies have direct oxygen-abundance measurements and which rely on UV line ratios or strong-line estimates. For GHZ2, we do not assign an O/H value and instead use only the UV line-ratio
relations for the abundance ratios.}
Tables \ref{table_abundance_stellar} and \ref{table_abundance_agn} summarize the gas-phase elemental abundance ratios for all of our galaxies.

\begin{deluxetable*}{lccc}
\tablecaption{\red{Electron Temperatures and Oxygen Abundance Diagnostics}}
\tablehead{
\colhead{ID} &
\colhead{$T_{\rm e}$({\sc [Oii]}) (K)} &
\colhead{$T_{\rm e}$ diagnostic} &
\colhead{O/H diagnostic} 
}
\startdata
GN-z11 & $<12328$ & {\sc Oiii]} / {\sc [Oiii]} $\lambda4363$$^{c}$ & direct, UV {\sc Oiii]} + {\sc [Oii]}  \\
GLASS 150008 & $18749^{+1852}_{-1356}$ & {\sc Oiii]} / {\sc [Oiii]}$^{b}$ & strong-line, $R_3$\\
CEERS 01019 & $15722^{+2649}_{-2801}$ & {\sc [Oiii]} $\lambda4363$$^{a}$ & direct, optical {\sc [Oiii]} + {\sc [Oii]}  \\
RXCJ2248-ID & $20300^{+2700}_{-2300}$ & {\sc [Oiii]} $\lambda4363$$^{a}$ & strong-line, $R_3$  \\
A1703-zd6 & $22955^{+7285}_{-5934}$ & {\sc [Oiii]} $\lambda4363$$^{a}$ & strong-line, $R_3$ \\
GN-z9p4 & $18225^{+5366}_{-5112}$ & {\sc [Oiii]} $\lambda4363$$^{a}$& direct, optical {\sc [Oiii]} +{\sc [Oii]} \\
GHZ2 & -- & assumed range & UV line-ratio relations only; no O/H \\
GHZ9 & $<19102$ &{\sc Oiii]} / {\sc [Oiii]} $\lambda4363$$^{c}$  & direct, UV {\sc Oiii]} + {\sc [Oii]}  \\
\enddata
\tablecomments{(1) ID. (2) Electron temperature of the O$^{++}$ zone.
Quoted uncertainties are the adopted asymmetric bootstrap errors where available. (3) Diagnostic used to derive $T_{\rm e}({\rm O^{++}})$. (4) Diagnostic used to estimate the oxygen abundance.
}
\tablenotetext{a}{Based on the auroral-to-nebular 
{\sc [Oiii]} $\lambda4363$/{\sc [Oiii]} $\lambda\lambda4959,5007$ ratio.}

\tablenotetext{b}{Based on the UV-to-optical 
{\sc Oiii]} $\lambda\lambda1661,1666$/{\sc [Oiii]} $\lambda\lambda4959,5007$ ratio.}

\tablenotetext{c}{Based on the UV-to-auroral ratio 
{\sc Oiii]} $\lambda\lambda1661,1666$/{\sc [Oiii]} $\lambda4363$.}

\label{table_Tene}
\end{deluxetable*}

\begin{deluxetable*}{l c cccccc c cccccc c}
    \rotate
    \tablecaption{Abundance Ratios Using Stellar ICF \label{table_abundance_icf}}
    
    \tabletypesize{\scriptsize}
    \tablewidth{0pt}
    \setlength{\tabcolsep}{2pt} 

    \tablehead{
        \colhead{} & \colhead{} & \multicolumn{6}{c}{UV Oxygen} & \colhead{} & \multicolumn{6}{c}{Optical Oxygen} & \colhead{} \\
        \cmidrule{3-8} \cmidrule{10-15}
        \colhead{ID} & \colhead{12+log(O/H)} & \colhead{[N/O]} & \colhead{[C/O]} & \colhead{[Ne/O]} & \colhead{[Ar/O]} & \colhead{[S/O]} & \colhead{[Fe/O]} & 
        \colhead{} & \colhead{[N/O]} & \colhead{[C/O]} & \colhead{[Ne/O]} & \colhead{[Ar/O]} & \colhead{[S/O]} & \colhead{[Fe/O]} & \colhead{[Ne/C]} \\
        \colhead{(1)} & \colhead{(2)} & \colhead{(3)} & \colhead{(4)} & \colhead{(5)} & \colhead{(6)} & \colhead{(7)} & \colhead{(8)} & 
        \colhead{} & \colhead{(9)} & \colhead{(10)} & \colhead{(11)} & \colhead{(12)} & \colhead{(13)} & \colhead{(14)} & \colhead{(15)}
    }

    \startdata
    GN-z11 & $ < 8.12 $ & $>0.64$ & $>-0.71$  & $ > 0.11$ & --$^a$ & --$^a$ & --$^a$ & &  --$^a$ & --$^a$ & --$^a$ & --$^a$ & --$^a$ & --$^a$ & $0.26^{+0.30}_{-0.48}$ \\
    GLASS\_150008 & $7.77 {\pm 0.03}$ & $0.54^{+0.30}_{-0.10}$ & $-0.54^{+0.40}_{-0.14}$ & --$^a$ & $<1.37$ & --$^a$ & $<0.32$ & &  $0.87^{+0.30}_{-0.10}$ & $-0.16^{+0.40}_{-0.14}$  & --$^a$ & $<1.38$ & --$^a$ & $<0.31$ & --$^a$ \\
    CEERS\_01019 & $7.94^{+0.46}_{-0.31}$ & $>1.25$ & --$^b$ & $>0.04$ & --$^b$ & --$^b$ &--$^b$ & & $1.11^{+0.42}_{-0.41}$ & $ <-0.02$  & $-0.04^{+0.01}_{-0.02}$ & $<2.55$ & $<0.72$ & --$^a$ & $>0.24$ \\ 
    RXCJ2248-ID & $7.64^{+0.03}_{-0.04}$ & $0.62^{+0.07}_{-0.05}$  & $-0.60^{+0.06}_{-0.07}$ & $0.15^{+0.01}_{-0.02}$ & $<3.00$ & $<0.46$ & $<0.25$ & & $0.48^{+0.08}_{-0.05}$ & $-0.62^{+0.05}_{-0.03}$  & $0.15^{+0.01}_{-0.02}$ & $<2.51$& $<0.51$ & $<0.34$ & $0.75 \pm 0.05$ \\ 
    A1703-zd6 & $7.53^{+0.05}_{-0.06} $ & $0.59^{+0.24}_{-0.25}$ & $-0.53^{+0.12}_{-0.13}$ & $0.07^{+0.01}_{-0.02}$ & $<2.19$ & --$^a$ & $<0.72$ & & $0.67^{+0.16}_{-0.15}$  & $-0.61^{+0.19}_{-0.23}$   & $0.05^{+0.02}_{-0.03}$ & $<2.19$ & --$^a$ & $<0.72$ & $0.61 \pm 0.10$ \\ 
    GN-z9p4 & $7.70^{+0.17}_{-0.13}$ & $ >0.78 $ & $ >-0.55 $ & $>0.40$ & --$^b$ & --$^a$ & --$^b$ & & $0.57 \pm 0.16$ & $-0.95 \pm 0.20$   & $-0.09 \pm 0.03$ & $<1.14$ & --$^a$ & $<1.23$ & $0.87 \pm 0.13$ \\
    GHZ2 & -- & $(0.19) - (0.23)$ & $(-0.91)-(-0.77)$ & $(-0.25)-(0.46)$ & --$^a$ & --$^a$ & --$^a$ &  & --$^a$ & --$^a$  & --$^a$ & --$^a$ & --$^a$ & --$^a$ & $0.95 \pm 0.46$ \\
    GHZ9 & $< 7.49$  &--$^a$ & --$^a$  & --$^a$ & --$^a$ & --$^a$ & --$^a$ & & $>0.63$ & $>-0.55$ & $>-0.15$ & --$^b$ & --$^b$ & --$^b$  & $0.01 \pm 0.32$ \\
    \enddata

    \tablecomments{
    Upper and lower limits correspond to the 3$\sigma$ flux upper bounds.\\
    $a$ The emission lines are out of the observed wavelength.\\
    $b$ Emission lines for both numerator and denominator are undetected.\\}
    \label{table_abundance_stellar}
\end{deluxetable*}

\begin{deluxetable*}{l c cccccc c cccccc c}
    \rotate
    \tablecaption{Abundance Ratios Using AGN ICF \label{table_abundance_agn}}

    \tabletypesize{\scriptsize}
    \tablewidth{0pt}
    \setlength{\tabcolsep}{2pt} 

    \tablehead{
        \colhead{} & \colhead{} & \multicolumn{6}{c}{UV Oxygen} & \colhead{} & \multicolumn{6}{c}{Optical Oxygen} & \colhead{} \\
        \cmidrule{3-8} \cmidrule{10-15}
        \colhead{ID} & \colhead{12+log(O/H)} & \colhead{[N/O]} & \colhead{[C/O]} & \colhead{[Ne/O]} & \colhead{[Ar/O]} & \colhead{[S/O]} & \colhead{[Fe/O]} & 
        \colhead{} & \colhead{[N/O]} & \colhead{[C/O]} & \colhead{[Ne/O]} & \colhead{[Ar/O]} & \colhead{[S/O]} & \colhead{[Fe/O]} & \colhead{[Ne/C]} \\
        \colhead{(1)} & \colhead{(2)} & \colhead{(3)} & \colhead{(4)} & \colhead{(5)} & \colhead{(6)} & \colhead{(7)} & \colhead{(8)} & 
        \colhead{} & \colhead{(9)} & \colhead{(10)} & \colhead{(11)} & \colhead{(12)} & \colhead{(13)} & \colhead{(14)} & \colhead{(15)}
    }

    \startdata
    GN-z11 & $ < 8.12 $ & $>0.62$ & $>-0.78$  & $>-0.12$ & --$^a$ & --$^a$ & --$^a$ & &  --$^a$ & --$^a$  & --$^a$ & --$^a$ & --$^a$ & --$^a$ & $0.13^{+0.28}_{-0.48}$ \\
    GLASS\_150008 & $7.77 {\pm 0.03}$ & $0.55^{+0.30}_{-0.10}$ & $-0.53^{+0.40}_{-0.14}$ & --$^a$ & $<1.01$ & --$^a$ & $<0.33$ & &  $0.89^{+0.30}_{-0.10}$ & $-0.12^{+0.10}_{-0.30}$ & $^a$ & $<0.94$ & --$^a$ & $<0.45$ & --$^a$ \\
    CEERS\_01019 & $7.94^{+0.46}_{-0.31}$ & $>1.29$ & --$^b$ & $-0.01^{+0.04}_{-0.03}$ & --$^b$ & --$^a$ & --$^b$ & & $1.23^{+0.43}_{-0.40}$ & $<-0.11$  & $-0.04^{+0.01}_{-0.02}$ & $<2.20$ & $<0.72$ & --$^a$ & $>0.32$ \\ 
    RXCJ2248-ID & $7.64^{+0.03}_{-0.04}$ & $0.70^{+0.07}_{-0.05}$  & $-0.10^{+0.06}_{-0.07}$ &  $0.02 \pm0.01$ & $<2.75$& $<0.38$ & $<0.39$ & & $0.15^{+0.08}_{-0.05}$ & $-0.48^{+0.05}_{-0.03}$  & $0.015 \pm0.01$ & $<2.16$ & $<0.32$ & $<0.70$ & $0.12 \pm 0.06$ \\ 
    A1703-zd6 & $7.53^{+0.05}_{-0.06} $ & $0.64 \pm0.15$ & $-0.54^{+0.12}_{-0.13}$ & $0.02^{+0.01}_{-0.02}$ & $<1.33$ & --$^a$ & $<0.53$ & & $0.63 \pm0.22$ & $-0.60^{+0.19}_{-0.23}$  & $-0.05^{+0.02}_{-0.04}$ & $<1.79$ & --$^a$ & $<0.68$ & $0.55 \pm 0.10$ \\ 
    GN-z9p4 & $7.70^{+0.17}_{-0.13}$ & $>0.79$ & $>-0.89$ & $>0.14$ & --$^b$ & --$^a$ & --$^b$ & & $0.57 \pm 0.16 $ & $-0.74 \pm 0.20 $ & $-0.10\pm 0.09$ &  $<1.51$ & --$^a$ &  $<1.24$ & $0.84 \pm 0.13$ \\
    GHZ2 & --& $(0.19) - (0.23)$ & $(-0.91)-(-0.77)$ & $(-0.25)-(0.46)$ & --$^a$ & --$^a$ & --$^a$ &  & --$^a$& --$^a$  & --$^a$ & --$^a$ & --$^a$ & --$^a$ & $0.95 \pm 0.46$ \\
    GHZ9 & $< 7.49$ & --$^a$ & --$^a$ & --$^a$ & --$^a$ & --$^a$ & --$^a$ & & $>0.63$ & $>-0.55$  & $>-0.15$ & --$^b$ & --$^b$ & --$^b$ & $-0.16 \pm 0.32$ \\
    \enddata

    \tablecomments{
    Upper and lower limits correspond to the 3$\sigma$ flux upper bounds.\\
    $a$ The emission lines are out of the observed wavelength.\\
    $b$ Emission lines for both numerator and denominator are undetected.}
\end{deluxetable*}

\section{Chemical Evolution Models}\label{sec:Nmodels}
To understand the origin of the high N/O, we compare the galactic chemical evolution models with the observations.
We construct the chemical evolution models of N/O, C/O, Ne/O, Ar/O, S/O and Fe/O ratios based on \cite{2024ApJ...962...50W}.
Our chemical evolution models are one-box models.
We assume instantaneous star formation based on the initial mass function of \cite{2001MNRAS.322..231K} for the fiducial conditions.
We also develop the models using the top-heavy IMF ($\alpha = 1.55$) of \cite{2012MNRAS.422.2246M}. 

We derive lifetimes of the stars as a function of masses from \cite{1998A&A...334..505P} and \cite{2018ApJ...857..111T}.
The ranges of time calculated by our models are between $2\times 10^{6}$ yr and $3\times10^{7}$ yr, which correspond to the lifetimes of 120 $M_\odot$ and 9 $M_\odot$ star, respectively.
The stars cause CCSN explosions after finishing their lifetimes.
Following the approach of \citet{2024ApJ...962...50W}, we develop chemical evolution models that incorporate the failed supernovae of massive stars. 
In these models, we assume that massive stars in the mass range of $15\text{--}120\,M_\odot$ undergo failed supernovae, a process proposed to occur in metal-poor environments. 
We adopt $15\,M_\odot$ as the transition mass; stars above this threshold collapse directly into black holes without a CCSN explosion, contributing no explosive nucleosynthesis products to the ISM.
Consequently, the chemical enrichment in our models is driven by only two channels: (1) nitrogen-rich material released through pre-collapse stellar winds (in the WR and SMS scenarios) or TDEs, and (2) CCSN ejecta from lower-mass stars in the $9\text{--}15\,M_\odot$ range.
At low metallicity ($Z=10^{-4}$), massive stars with masses $\gtrsim 25~M_\odot$ may collapse directly into black holes because they form massive Fe cores at the final stage of their evolution \citep{2020ApJ...888...91E}. 
Following the yields of \citet{N13papaer}, we sum the ejecta from CCSNe in the $9\text{--}15\,M_\odot$ range to derive the chemical abundance ratios of our models.

To explain the abundance ratios in our high-$z$ galaxies, we focus on three nitrogen-enrichment scenarios: WR stars, SMSs, and TDEs.
We develop a chemical evolution model that incorporates the effects of each of these three scenarios.

In the WR models, we adopt metallicity values of [Fe/H] = $-3$  (fiducial) and $-1$ from the WR yields of \cite{2018ApJS..237...13L} to represent the low-metallicity environments characteristic of high-redshift galaxies. 
We select a rotation velocity of $300~\mathrm{km/s}$, as this velocity maximizes the resulting N/O ratios.
The rotation of WR stars induces the mixing of the outer hydrogen and helium layers.
Unlike non-rotating stars where nitrogen enrichment is limited to secondary production, rotating massive stars can produce primary nitrogen. This is achieved by the mixing of C and O from the helium-burning core into the hydrogen-burning shell, where they are converted into nitrogen via the CNO cycle before being transported to the surface.
During the WR stages, the WR stars with $25-120~M_\odot$ eject the surface material rich in nitrogen via the stellar winds.
The stars with $9\text{--}15\,M_\odot$ undergo CCSNe without passing through the WR phase.
For these stars, we adopt the CCSN yields from \cite{N13papaer}, which originally cover $13\text{--}40\,M_\odot$. 
The yields for the $9\text{--}13\,M_\odot$ range are obtained by extrapolating the \cite{N13papaer} data. 
We add up the ejecta from both stellar winds ($25-120~M_\odot$) and CCSNe ($9-15~M_\odot$) along the period of the WR stage and the stellar lifetime.

In the SMS models, we utilize the yields of $10^5~M_\odot$ \citep{2023ApJ...949L..16N}, because a SMS with $10^5~M_\odot$ produces the highest N/O in their yields.
In the SMS, the CNO cycle is activated due to the inflow of the pristine gas \citep{2018MNRAS.478.2461G}.
The stellar winds from the SMS eject a large amount of nitrogen. 
\cite{2023ApJ...949L..16N} assume that the metallicities of SMS yields are 0.1$\mathrm{Z_\odot}$.
In the SMS model, we assume the presence of a single SMS coexisting with a population of stars with masses up to 100 M$_{\odot}$ distributed according to an IMF.
We add up the ejecta from the stellar wind of a SMS and CCSNe of $9-15~M_\odot$.
We adjust the mass ratio between a single SMS and stars with masses $< 100~M_\odot$ such that their combined mass is constrained to a total of $10^8~M_\odot$.

\begin{figure*}
    \centering
    \includegraphics[width=18cm]{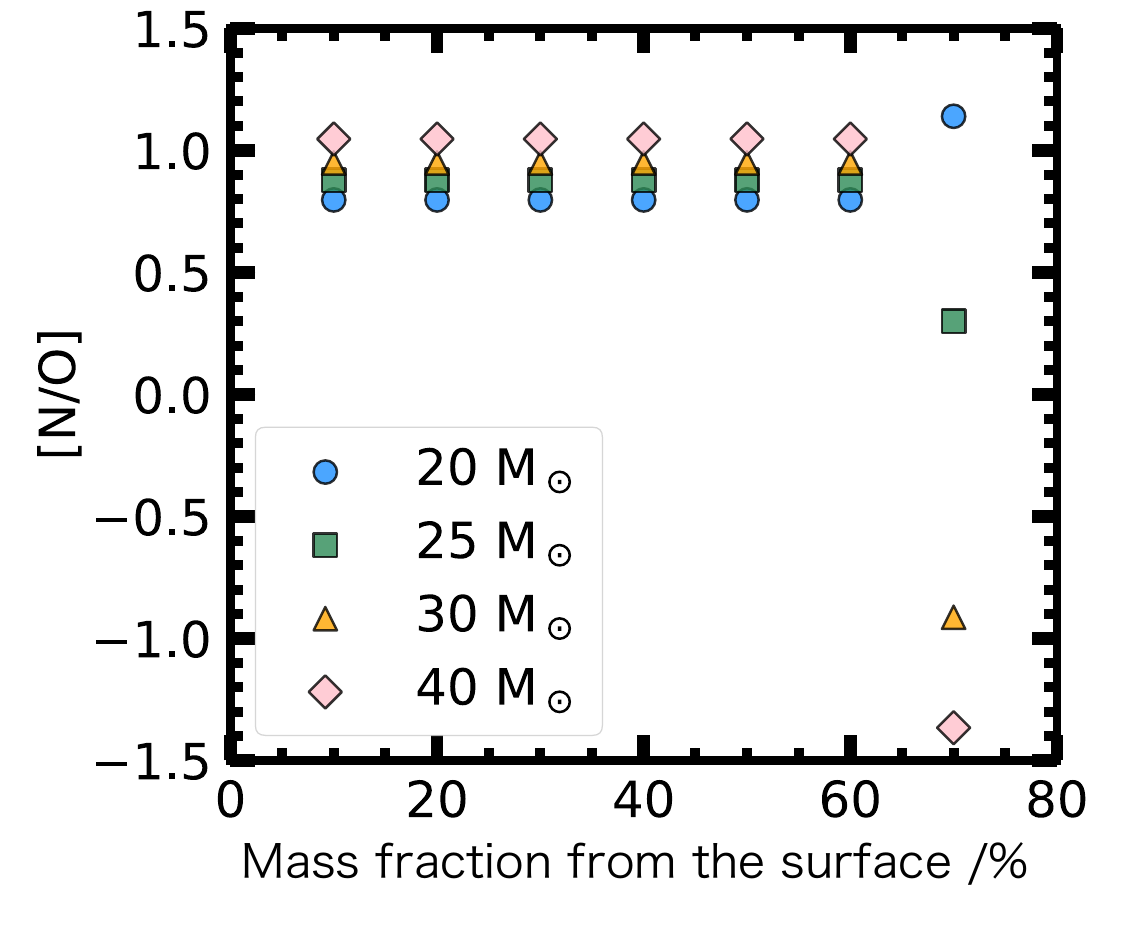}
    \caption{[N/O] abundance ratio in the yields of TDE as a function of the mass fraction stripped from the stellar surface. Different symbols represent stars with initial masses of 20 (blue circles), 25 (green squares), 30 (orange triangles), and 40 M$_\odot$ (pink diamonds).}
    \label{fig:TDE}
\end{figure*}

In the TDE models, we assume that the outer H layers of the stars with $9-100~M_\odot$ are disrupted by the tidal force of a black hole.
The stellar envelope is disrupted by the tidal force of the black hole and subsequently released into the ISM.
Similar to WR winds, this releases the nitrogen-rich material from the stellar surface.
Because detailed TDE yields are not available in the literature, we have calculated them using the nucleosynthesis code from \cite{2007ApJ...660..516T}. 
In these calculations, we assume that the stellar hydrogen-rich envelope mass in a star's pre-supernova is ejected during the TDE because the production of a significant amount of nitrogen via the CNO cycle requires sufficient time,  reaching its peak by the pre-supernova stage.
Figure \ref{fig:TDE} shows [N/O] as a function of the mass fraction stripped from the surface in our TDE yields for stars with initial masses of 20–40 $M_\odot$. In our TDE yields, the [N/O] ratio remains nearly constant when 10–60\% of the stellar mass is stripped, whereas it decreases significantly for most masses when the stripped fraction reaches 70\%.
In our model, we assume the TDE of the outer 50\% of the stellar mass. 
While disruptions exceeding 60\% of the stellar mass lead to an increase in oxygen abundance from the deeper layers, a 50\% disruption efficiently incorporates the nitrogen-rich H-envelope while maintaining a high N/O ratio by avoiding significant oxygen contamination.
The stellar metallicities of TDE yields are $Z = 0.001$ (fiducial) and $0.1$ in order to represent the low-metallicity environments characteristic of high-redshift galaxies.
The total abundance ratios in this scenario are the sum of the material ejected from TDEs and the yields from subsequent CCSN explosions within the same stellar population.
Stars undergoing TDEs release nitrogen-rich envelopes into the ISM without undergoing a CCSN. 
These models yield a higher nitrogen abundance compared to standard CCSN models.

We summarize our models of the WR, SMS, and TDE in Table \ref{table_nitrogenmodel}.

\begin{deluxetable*}{ccccc}
    \tabletypesize{\scriptsize}
    \tablewidth{0pt} 
    \tablecaption{Models of the Nitrogen Origin \label{table_nitrogenmodel}}
    \tablehead{\colhead{Model} & \colhead{nitrogen origin}& \colhead{Progenitor Star} & \colhead{ Yields } & \colhead{Reference}
    \\
    (1) &(2) &(3) & (4) & (5)}
     \startdata 
     {   }& {    } & {$9~ M_\odot \leq  M  < 15~M_\odot$} & {CCSN}  &  {\cite{N13papaer}}  \\ 
     {WR Model } & {WR stars} & { $15 ~ M_\odot \leq  M \leq 120~M_\odot$ } & {failed supernovae} & { }  \\ 
     {  } & { } &{$25 ~ M_\odot <  M \leq 120~M_\odot$}&  {WR stellar wind} & {\cite{2018ApJS..237...13L}} \\
      \hline
     { } & { } &{$9~ M_\odot \leq  M  < 15~M_\odot$} & {CCSN}  &  {\cite{N13papaer}} \\
      {SMS Model } & { SMS } & { $15 ~ M_\odot \leq  M \leq 120~M_\odot$ } & {failed supernovae} & { }\\
      {   }& {    } & {$10^5~M_\odot$} & {SMS stellar wind}  &  {\cite{2023ApJ...949L..16N}}  \\ 
    \hline
     {   }& {    }& {$9~ M_\odot \leq  M  < 15~M_\odot$} & {CCSN}  &  {\cite{N13papaer}}  \\ 
     {TDE Model } & {TDE} & { $15 ~ M_\odot <  M \leq 100~M_\odot$ } & {failed supernovae} & { }  \\ 
     { } & { } & { $9 ~ M_\odot \leq  M \leq 100~M_\odot$} & {TDE} & This paper \\
    \enddata
    \label{table_nitrogenmodel}
\end{deluxetable*}
\section{Results \& Discussion} \label{sec:result}

\subsection{N/O ratios vs. Other Abundance Ratios}\label{sec:result_2}

\begin{figure*}
    \centering
    \includegraphics[width=16cm]{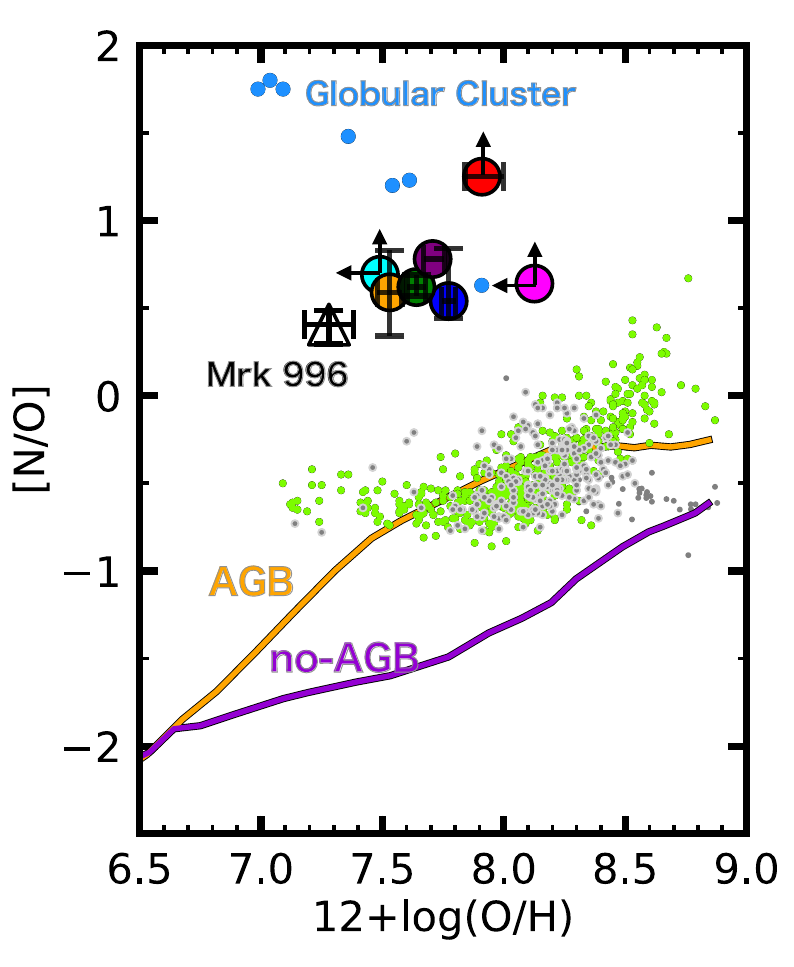}
    \caption{\red{
    The nitrogen-to-oxygen abundance ratio as a function of oxygen abundance.
    The large filled circles, colored by target (magenta: GN-z11, blue: GLASS\_150008, red: CEERS\_01019, green: RXCJ2248-ID, orange: A1703-zd6, purple: GN-z9p4, and cyan: GHZ9), represent the abundance ratios derived using stellar ICFs and UV oxygen lines.
    GHZ2 is not shown because its oxygen abundance cannot be determined from the available emission lines.
    The black open symbol represents Mrk 996.
    The blue circles represent dwarf turnoff stars in the globular cluster NGC 6752 \citep{2005A&A...433..597C}, while the green circles show $z\sim0$ {\sc Hii} regions \citep{2012MNRAS.424.2316P,2020ApJ...893...96B}.
    The gray circles denote local dwarf galaxies \citep{2006A&A...448..955I,2019ApJ...874...93B}.
    The orange and purple curves indicate the chemical evolution models including CCSNe, HNe, and SNe Ia with and without AGB-star enrichment, respectively \citep{N13papaer}.
    Our galaxies lie well above the locus of normal star-forming galaxies at fixed $12+\log({\rm O/H})$, confirming their enhanced nitrogen abundances.
    }}
    \label{fig:NOvsOH}
\end{figure*}

\begin{figure*}
    \centering
    \includegraphics[width=18cm]{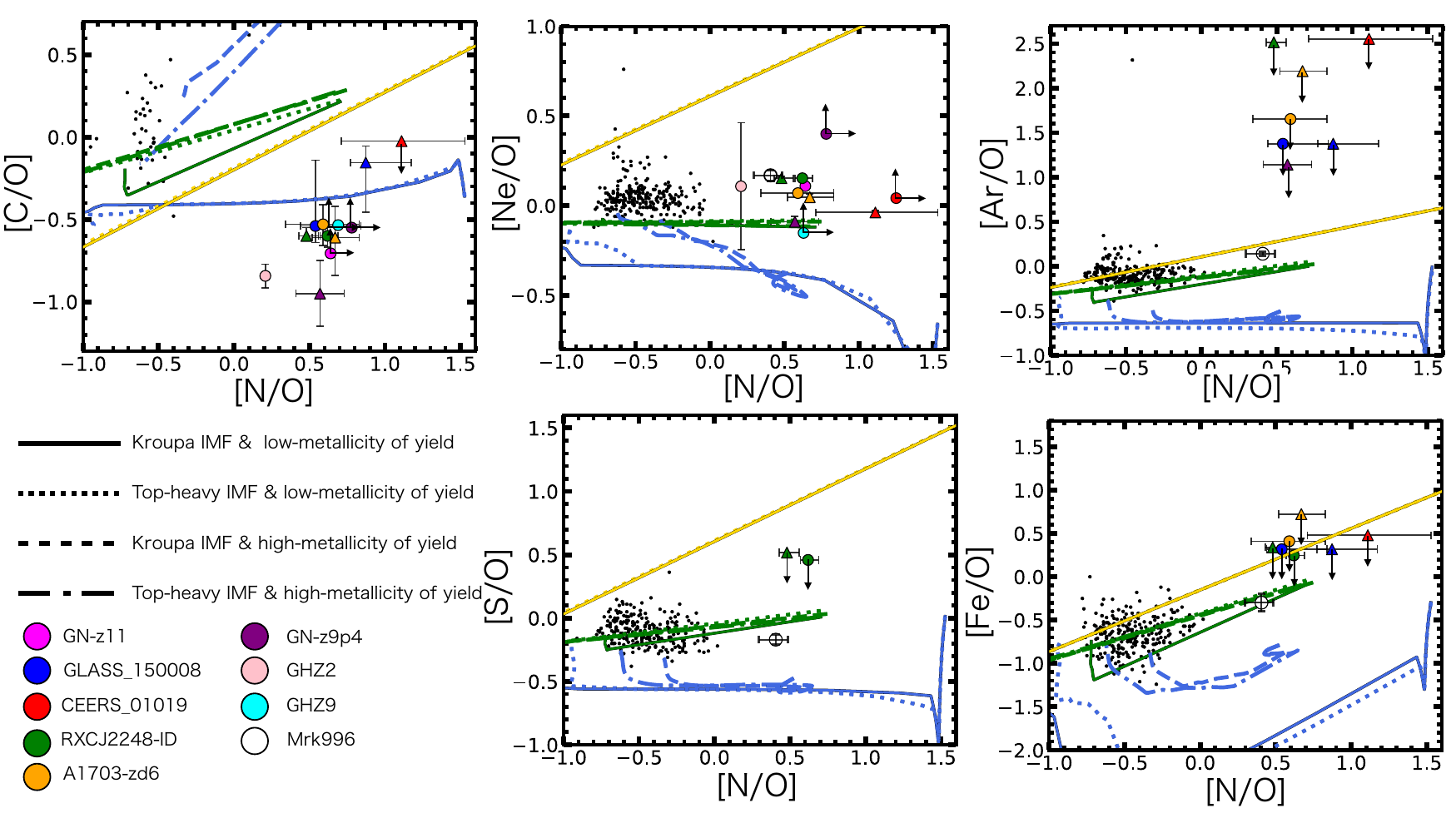}
    \caption{\red{
    Comparison between the observed abundance ratios and our chemical evolution models.
    The filled circles and triangles show the abundance ratios derived with stellar ICFs using UV and optical oxygen lines, respectively.
    Symbols are colored by target: magenta for GN-z11, blue for GLASS\_150008, red for CEERS\_01019, green for RXCJ2248-ID, orange for A1703-zd6, purple for GN-z9p4, pink for GHZ2, and cyan for GHZ9.
    The black dots show local dwarf galaxies \citep{2006A&A...448..955I,2019ApJ...874...93B}, and the open circles show Mrk 996.
    The blue, yellow, and green curves represent the WR, SMS, and TDE models, respectively.
    The solid curves show the fiducial models with low-metallicity yields and a Kroupa IMF.
    The dotted curves show models with low-metallicity yields and a top-heavy IMF.
    The dashed and dash-dotted curves show models with high-metallicity yields and a Kroupa and top-heavy IMF, respectively.
    }}
    \label{fig:NOvsArO}
\end{figure*}

\begin{figure*}
    \centering
    \includegraphics[width=18cm]{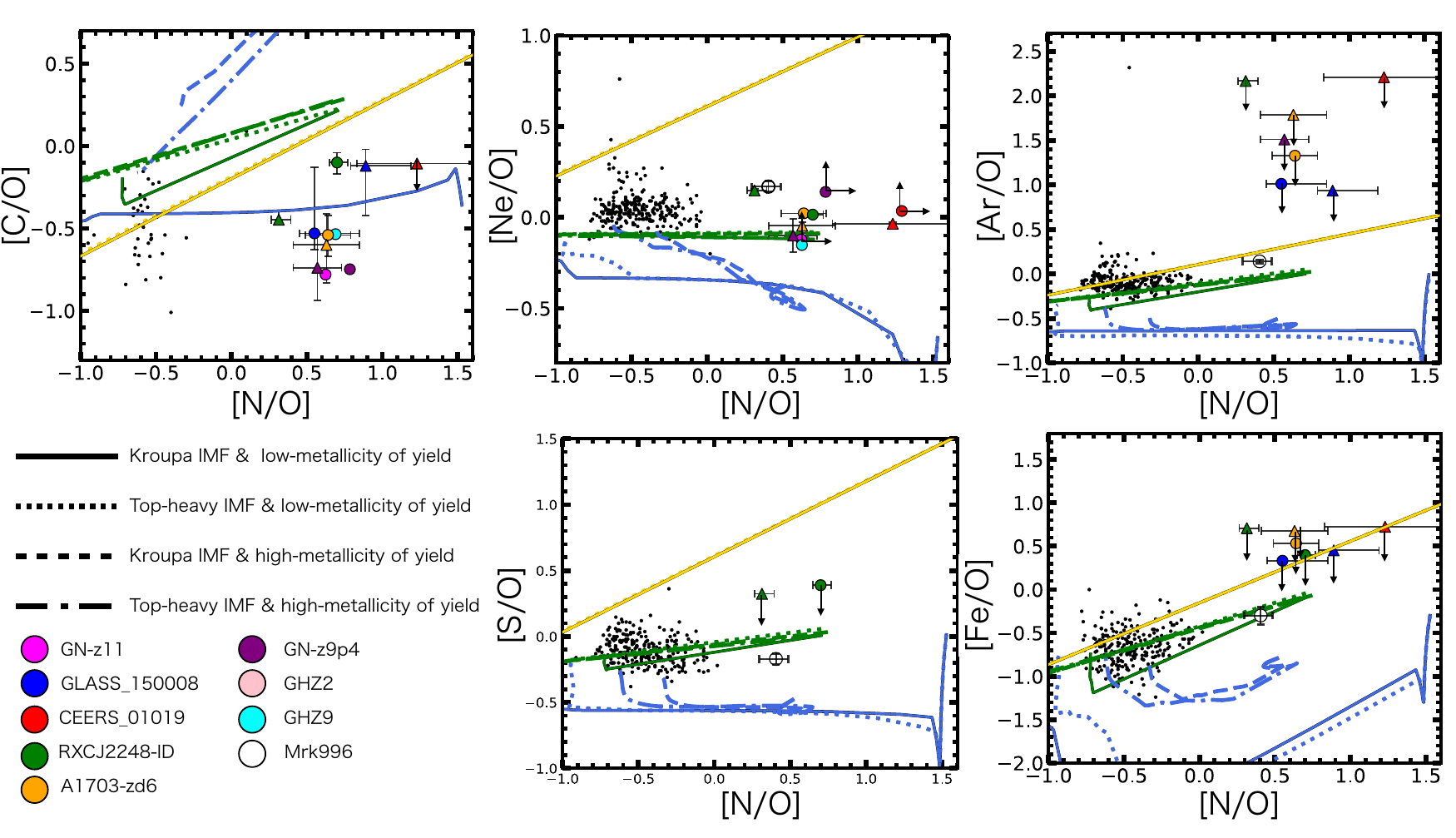}
    \caption{\red{
    Same as Figure~\ref{fig:NOvsArO}, but for abundance ratios derived with AGN ICFs.
    }}
    \label{fig:NOvsArO_agn}
\end{figure*}

\begin{figure*}
    \centering
    \includegraphics[width=18cm]{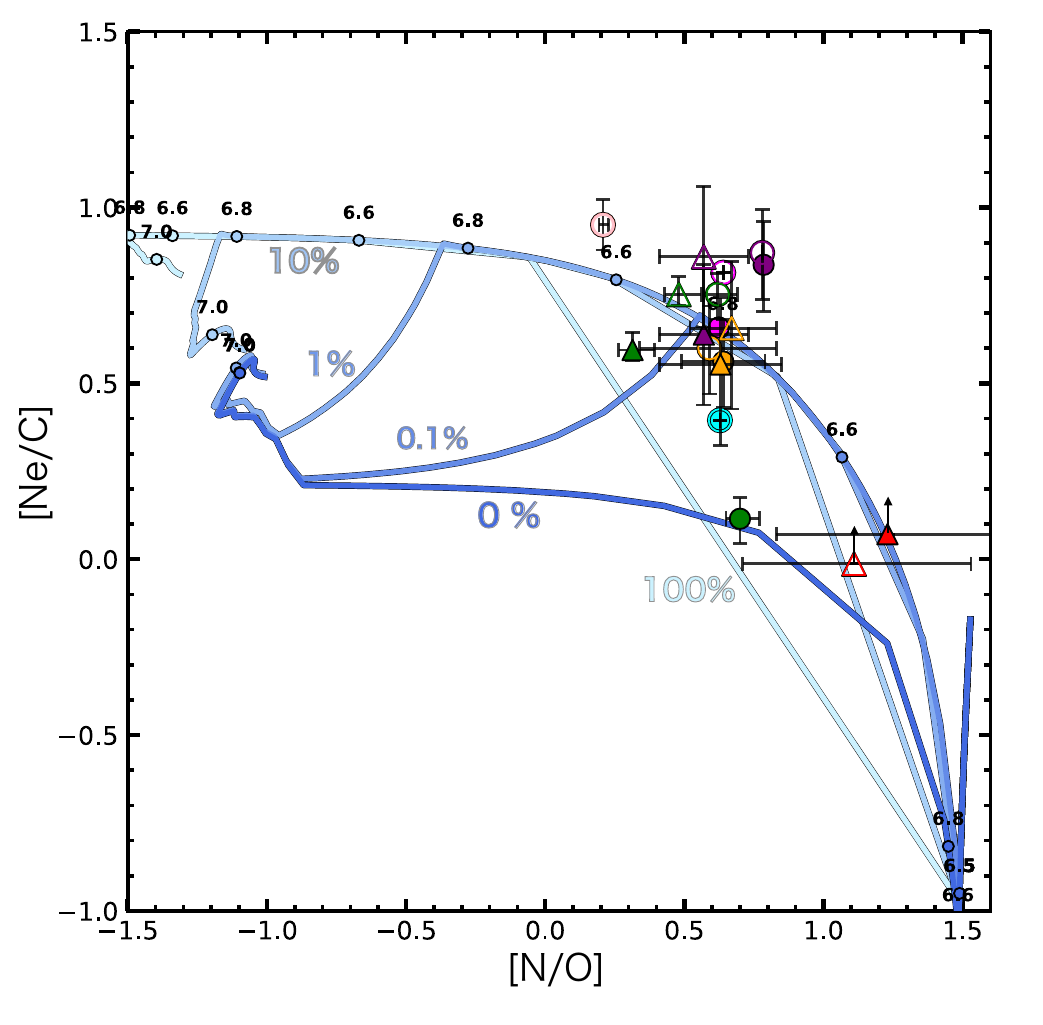}
    \caption{\red{
    Comparison between the observed [Ne/C] and [N/O] abundance ratios and the WR chemical evolution models.
    The filled circles and triangles show the abundance ratios derived with stellar ICFs using UV and optical oxygen lines, respectively.
    The corresponding open symbols show the results derived with AGN ICFs.
    Symbols are colored by target as in Figure~\ref{fig:NOvsArO}.
    The curves show WR models with [Fe/H]$=-3$ and a Kroupa IMF, with different fractions of massive stars contributing CCSN ejecta.
    The CCSN fractions are 0, 0.1, 1, 10, and 100\%, as indicated by the curve colors.
    The small circles and numerical labels along the model tracks indicate the evolutionary ages in $\log(t/{\rm yr})$.
    }}
    \label{fig:NCvsNO}
\end{figure*}

\begin{figure*}
    \centering
    \includegraphics[width=18cm]{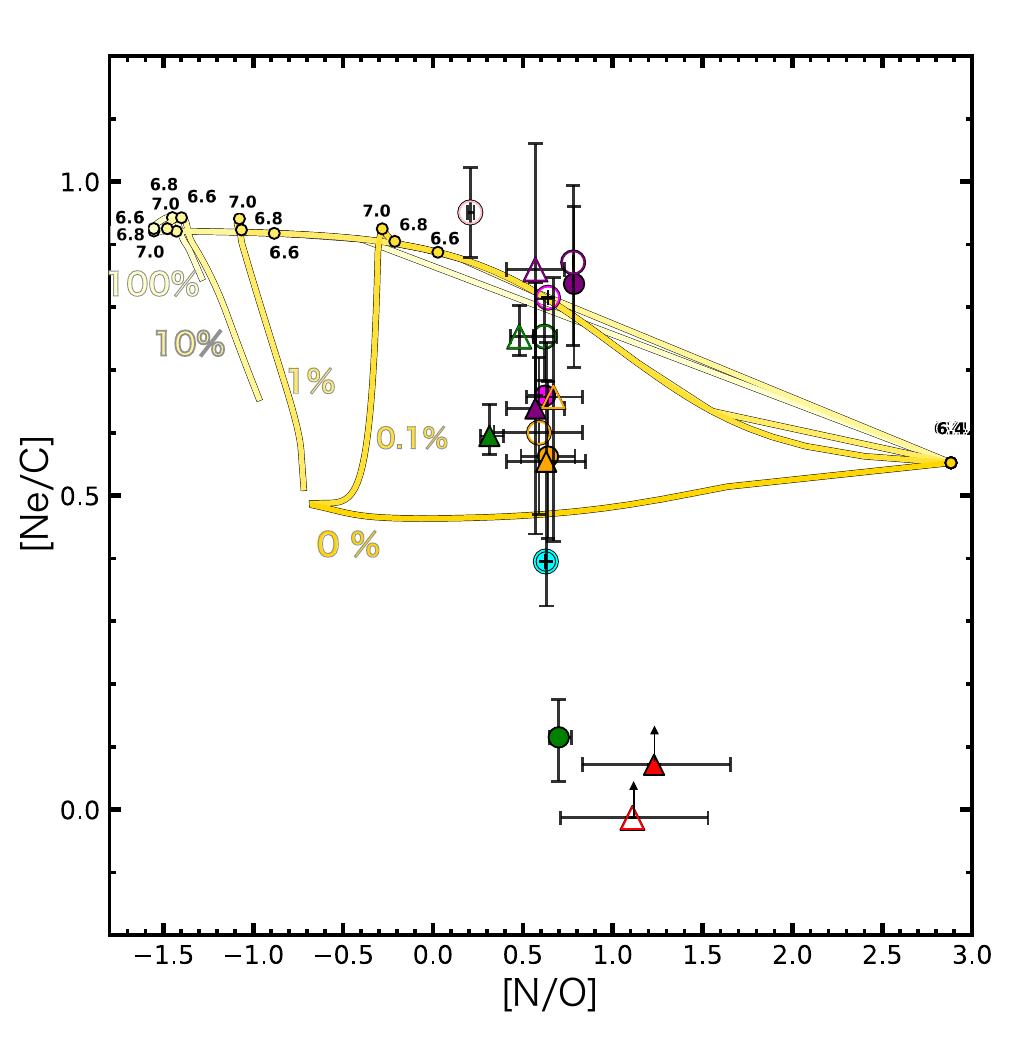}
    \caption{Same as Figure \ref{fig:NCvsNO}, but for the SMS models.}
    \label{fig:NCvsNO_SMS}
\end{figure*}

\begin{figure*}
    \centering
    \includegraphics[width=18cm]{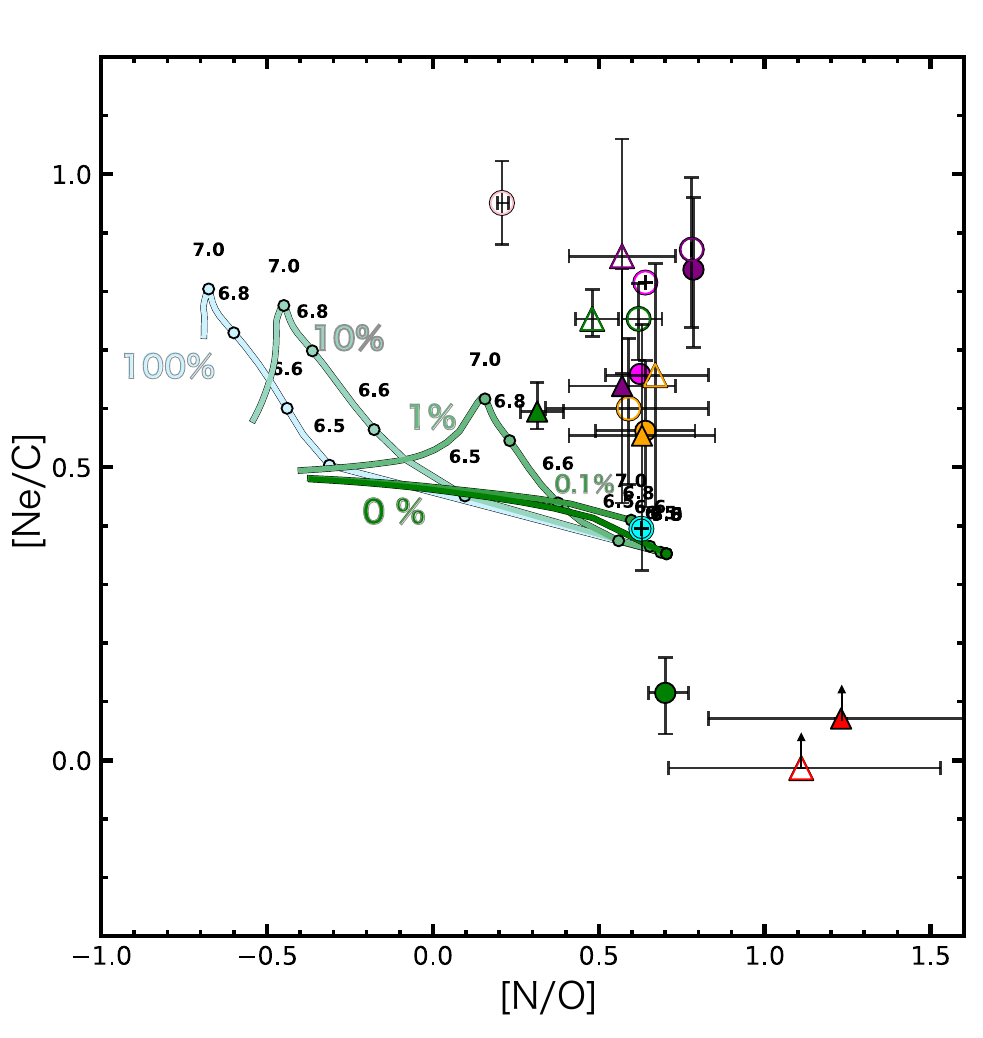}
    \caption{Same as Figure \ref{fig:NCvsNO}, but for the TDE models.}
    \label{fig:NCvsNO_TDE}
\end{figure*}

\red{Figure~\ref{fig:NOvsOH} shows [N/O] as a function of $12+\log({\rm O/H})$ for our sample and comparison galaxies.
GHZ2 is not included in this figure because its oxygen abundance cannot be
determined from the available emission lines.
Our galaxies are located well above the locus of local star-forming galaxies at fixed oxygen abundance, confirming that they are strongly enhanced in nitrogen relative to normal galaxies. 
Although AGB stars can increase N/O at later times, the AGB model based on the yields of \citet{N13papaer} remains below the observed values for our sample.
This indicates that standard AGB enrichment alone cannot explain the nitrogen excess in our sample.}

Figures \ref{fig:NOvsArO} and \ref{fig:NOvsArO_agn} show the abundance ratios for our high-$z$ galaxies derived using two different ICFs: those assuming stellar radiation and those assuming AGN radiation. 
Although there are minor discrepancies between the two sets of ICFs, the overall trends remain unchanged regardless of the assumed ionizing radiation field.
We also compare results based on UV {\sc Oiii]} lines (circles) and optical {\sc [Oiii]} lines (triangles). 
No significant differences are found between the abundance ratios derived from UV oxygen lines ({\sc Oiii]}$\lambda\lambda1661,1666$) and those from optical oxygen emission lines ({\sc [Oiii]}$\lambda\lambda4959,5007$)

Figures \ref{fig:NOvsArO} and \ref{fig:NOvsArO_agn} also show the abundance ratios of local dwarf galaxies \citep{2006A&A...448..955I, 2019ApJ...874...93B} for comparison.
The [N/O] ratios of our high-$z$ galaxies are significantly higher than those of local dwarf galaxies, while their [C/O] ratios are lower.
The other abundance ratios, [Ne/O], [Ar/O], [S/O], and [Fe/O] are consistent with those of the local dwarf galaxies, suggesting that only nitrogen is preferentially enriched. 

We also plot Mrk 996, a known WR galaxy with a high [N/O] ratio, as an open circle in Figures \ref{fig:NOvsArO} and \ref{fig:NOvsArO_agn}.
Following the method described in Section \ref{sec:abundance}, we derive the abundance ratios for Mrk 996 using the flux values from \cite{2011ApJ...734...82I}. 
The [C/O] ratio of Mrk 996 is not determined because the carbon emission lines fall outside the observed wavelength range.
In contrast to the local dwarf galaxies, the abundance ratios of Mrk 996 are similar to those of our high-$z$ galaxies. 
Mrk 996 is characterized not only by a high [N/O] ratio but also by the enrichment of other elements relative to oxygen in Figures \ref{fig:NOvsArO} and \ref{fig:NOvsArO_agn}. 
However, elements such as Ne, Ar, S, and Fe are synthesized deep within stellar interiors and are not typically carried away by WR stellar winds. This discrepancy suggests that a source of chemical enrichment beyond the standard WR scenario is required for Mrk 996. 
Among the models discussed in this work, a TDE provides the most compelling explanation for the observed abundance patterns in Mrk 996.

In Figures \ref{fig:NOvsArO} and \ref{fig:NOvsArO_agn}, we compare our observational data with our models for the WR, SMS, and TDE scenarios calculated for a range of IMFs and metallicities (Section \ref{sec:Nmodels}).
We use different line styles to represent the various assumptions for our models.
Our WR fiducial models assume a Kroupa IMF and a metallicity of [Fe/H] = -3.
We also show the low metallicity WR models of Kroupa IMF (solid) and Top-heavy IMF (dotted), respectively. 
Because there is little to no difference between our WR models based on the Kroupa IMF and Top-heavy IMF, our models show no significant dependence on the IMF.
This is because the predicted abundance ratios of our models in the integrated ejecta are largely insensitive to the underlying stellar mass distribution.
The abundance ratios of our WR models exhibit a clear dependence on the metallicity of the progenitor star.

Our TDE fiducial models assume a Kroupa IMF and a metallicity of $Z=0.001$.
Our TDE models show a similar trend to our WR models described above.
This metallicity effect is even stronger in our WR models than in our TDE models because the nucleosynthetic yields of WR stars are highly sensitive to their initial heavy-element content.
Note that our SMS models (yellow) are calculated only for a single metallicity (0.1$Z_\odot$), thus only the solid and dash-dotted curves are shown to show the IMF dependence.

Because the mass loss from the SMS stellar wind dominates the total chemical yield, the variations in the IMF of the underlying stellar population are masked and become negligible in the resulting abundance ratios.
In our SMS model, the yields of SMS assume an initial metallicity of 10\% solar. 
During stellar evolution, the CNO cycle efficiently converts oxygen into nitrogen, leading to a decrease in the O/H ratio and a corresponding increase in the N/O ratio. 
This oxygen depletion subsequently elevates the abundance ratios of other elements relative to oxygen (e.g., [Ne/O], [Ar/O], [S/O], and [Fe/O]), even in the absence of additional nucleosynthesis for those species.

The [C/O] ratios observed in our galaxy sample are either comparable to or lower than the values predicted by the WR models in Figures \ref{fig:NOvsArO} and \ref{fig:NOvsArO_agn}.
The equilibrium values of the CNO cycle are [N/O] $\sim 1.96$--$2.06$ and [C/O] $\sim -0.25$--$0.0$ \citep{2015A&A...576A..56M}.
While \cite{2023ApJ...959..100I} suggested that their sample galaxies exhibited ratios close to these CNO equilibrium values, our galaxies including the reanalyzed data from \cite{2023ApJ...959..100I} show [N/O] and [C/O] ratios that are lower than the equilibrium state.
This suggests that oxygen is more abundant than predicted by CNO equilibrium alone. 
Therefore, additional oxygen from sources such as CCSNe likely pollutes the CNO-processed gas.
The measured [Ne/O] ratios exceed the WR model predictions, indicating that no single enrichment scenario can simultaneously reproduce both the observed [C/O] and [Ne/O] ratios. 
Because Ne is synthesized deep within stellar interiors (e.g., the ONeMg core), it is not significantly ejected via WR stellar winds. 
The observed [Ne/O] ratios cannot be fully explained by the WR scenario alone.
To resolve this discrepancy, a scenario is required that enhances the Ne/O ratio while simultaneously suppressing the [C/O] ratio, all while maintaining the observed high [N/O] ratio. 
Increasing Ne abundance is typically challenging because it is co-produced with oxygen in the same stellar cores. 
Any increase in Ne is often accompanied by a rise in oxygen. 
This subsequent oxygen enrichment then leads to a decrease in the [N/O] ratio.
To address this, we develop chemical evolution models by injecting a trace amount of CCSN ejecta into our three models to provide the required neon.
This approach adds oxygen only to an extent that avoids decreasing the [N/O] ratio, maintaining the observed nitrogen enrichment.
Figures \ref{fig:NCvsNO}-\ref{fig:NCvsNO_TDE} show the model tracks for various CCSN fractions ($f_{\rm CCSN} = 0, 0.1, 1, 10,$ and $100\%$) for stars with $M \gtrsim 25\,M_\odot$. 
In Figure \ref{fig:NCvsNO}, we find that WR models with $f_{\rm CCSN} = 0\%$ and $100\%$ fail to reproduce the observed [Ne/C] ratios, yielding values lower than the observations.
In contrast, WR models with $f_{\rm CCSN} = 0.1, 1,$ and $10\%$ are consistent with the observed [Ne/C] range.
This is because the pure WR models ($f_{\rm CCSN}=0 \% $) predict neon abundances significantly lower than those observed in our galaxies. 
By incorporating a trace amount of CCSN ejecta, the model provides the required neon while keeping oxygen enrichment low enough to preserve the high [N/O] ratio.
The evolutionary ages, also indicated in Figure \ref{fig:NCvsNO}, provide further constraints.
For the WR models with $f_{\rm CCSN} = 1$ and $10\%$, the [N/O] ratios reach $-0.5$ and $0.5$ by $10^{6.6}$~yr, respectively, and the models are consistent with the observed abundances only over a narrow time window (e.g., $10^{6.5}$--$10^{6.6}$~yr for $f_{\rm CCSN} = 10\%$). In contrast, the WR model with $f_{\rm CCSN} = 0.1\%$ maintains consistency with the observations over $10^{6.6}$--$10^{6.8}$~yr, a factor of $\sim$2 longer in logarithmic timescale.
Figure \ref{fig:NCvsNO_SMS} illustrates the SMS models combined with varying CCSN fractions ($f_{\rm CCSN} = 0, 0.1, 1, 10,$ and $100\%$). 
For the majority of our galaxies, the observed [Ne/C] ratios cannot be reconciled with the SMS models at any $f_{\rm CCSN}$. 
Although some galaxies (purple, magenta, and pink objects) overlap with the SMS models across various $f_{\rm CCSN}$ values, this consistency lasts for only about 1.5 Myr. 
Such a short duration makes the SMS scenario an unlikely explanation for the observed population.
The TDE models with additional CCSN components are shown in Figure \ref{fig:NCvsNO_TDE}. 
As previously established in Figures \ref{fig:NOvsArO} and \ref{fig:NOvsArO_agn}, the TDE scenario inherently predicts [C/O] ratios that are higher than those measured in our galaxies. 
This initial carbon over-abundance implies that, even with the supplementary addition of CCSNe, the resulting [Ne/C] ratios remain lower than our observations.
This result suggests a physical scenario where the vast majority of massive stars ($M \gtrsim 25\,M_\odot$) undergo failed supernovae, effectively preventing the over-enrichment of oxygen. 
Only a tiny fraction ($0.1\%$) of such stars explode as CCSNe, providing the necessary amount of neon to match the observed [Ne/C] ratios without significantly lowering the [N/O] ratios.
We conclude that the $0.1\%$ CCSN fraction model best accounts for the simultaneous constraints of the N/O ratio, the Ne/C ratio, and the stellar age.

For the [S/O], [Ar/O], and [Fe/O] ratios in Figures \ref{fig:NOvsArO} and \ref{fig:NOvsArO_agn}, the derived upper limits from our high-$z$ galaxies are not enough to provide definitive constraints on the models. 
The predicted yields of these elements are sensitive to the physical conditions in deep stellar interiors. 
[S/O], [Ar/O], and [Fe/O] ratios are sensitive to the explosive nucleosynthesis conditions in CCSNe, enabling tighter constraints on the CCSN contribution fraction.
Therefore, measurements of elements synthesized in these layers are crucial for differentiating between the models.
While the jet-driven supernova models introduced by \citet{2024ApJ...974..310L} do not produce significant nitrogen yields, they exhibit [Ne/O] ratios of $0.11$--$0.14$, which are closely aligned with our sample. Furthermore, these models show [Ar/O], [S/O], and [Fe/O] ratios consistent with our observations, suggesting that jet-driven supernovae could potentially contribute to the chemical enrichment of these systems.

\subsection{N/O Values vs. Stellar Age} \label{sec:result_1}
\begin{figure*}
    \centering
    \includegraphics[width=18cm]{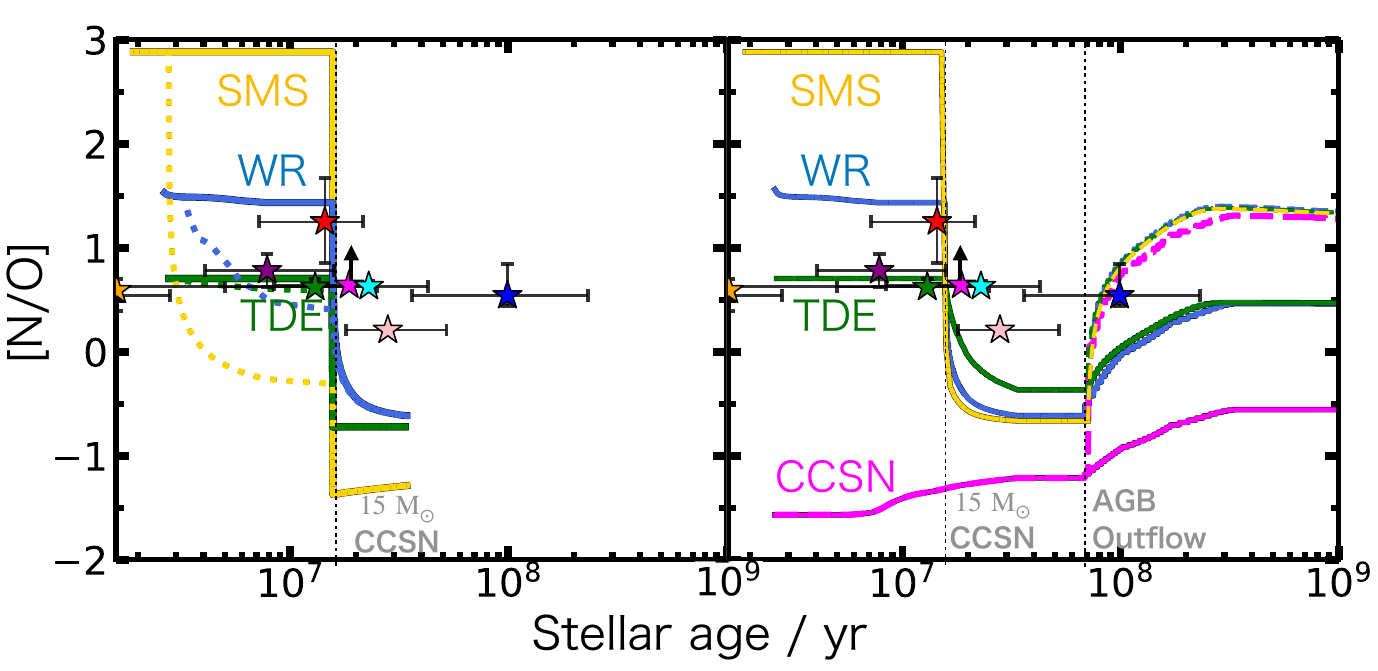}
    \caption{Comparison of our galaxies with the N/O ratios of our models as a function of time.
    The star symbols represent our galaxies: GN-z11 (magenta), GLASS\_150008 (blue), CEERS\_01019 (red), RXCJ2248-ID (green), A1703-zd6 (orange), GN-z9p4 (purple), GHZ2 (pink), and GHZ9 (cyan).
    The N/O ratios of our galaxies are derived using UV oxygen emission lines and stellar ICFs.
    The stellar ages of our galaxies derived by SED fitting are taken from \cite{Bunker_2023}, \cite{2023ApJ...951L..17J}, \cite{2024A&A...681A..30M}, \cite{2024MNRAS.529.3301T}, \cite{2025ApJ...980..225T}, \cite{Schaerer_2024}, \cite{2024ApJ...972..143C}, and \cite{2025A&A...693A..50N}.
    The left panel presents our failed supernova models. 
    The blue, yellow, and green curves show our WR, SMS, and TDE models, respectively.
    The magenta curve in the left panel represents the CCSNe models.
    The solid curve indicates a scenario where all stars with $M \gtrsim 15\,M_\odot$ undergo failed supernovae, while the dotted curve represents a model where 0.1\% of stars in this mass range explode as CCSNe.
    The right panel shows models incorporating AGB stars and outflows into the WR, TDE, and SMS scenarios. 
    The solid curves denote the three scenarios including the AGB contribution. 
    The dashed curves represent the outflow models, where 90\% of the CCSNe ejecta is effectively removed from the system.
    The vertical line at approximately $2\times 10^{7}$ yr marks the timing of CCSNe from $15\,M_\odot$ stars. Additionally, the vertical line near $7\times10^{7}$ yr indicates the onset of the AGB star contribution and the timing of the outflows.
}
    \label{fig:NOvsage}
\end{figure*}

The left panel of Figure \ref{fig:NOvsage} presents the N/O ratio as a function of stellar age for our measurements (Section \ref{sec:abundance}) and our models (Section \ref{sec:Nmodels}).
In Figure \ref{fig:NOvsage}, the stellar ages are adopted from the SED fitting results reported in previous studies \citep{Bunker_2023, 2023ApJ...951L..17J, 2024A&A...681A..30M, 2024MNRAS.529.3301T, 2025ApJ...980..225T, Schaerer_2024, 2024ApJ...972..143C, 2023ApJ...948L..14C}. 
For galaxies where the stellar age is not provided (GN-z9p4 and GHZ9), we estimate the stellar age by dividing the stellar mass ($M_*$) by the star formation rate, assuming a constant star formation history.
Our three models of the WR, SMS, and TDE (the blue, yellow and green curves) predict high N/O ratios of [N/O] = 1.5, 3, and 0.5, respectively.
In the left panel of Figure \ref{fig:NOvsage}, our models for the WR, SMS, and TDE scenarios are presented, considering two cases for stars with $M \gtrsim 15\,M_\odot$: one where all stars undergo failed supernovae, and another where 0.1\% of these stars explode as CCSNe.
These predicted values of our models are consistent with the N/O ratios (or their lower limits) observed in our high-$z$ galaxies.
Without failed supernovae, the massive oxygen release from CCSNe would rapidly decrease the N/O ratio within $\sim 10^{6.6}$~yr, making it impossible to sustain the elevated N/O ratios over the stellar ages observed in our high-$z$ galaxies.
\red{This requirement is mainly a timescale constraint. Nitrogen-rich material can raise N/O at early times in the WR, SMS, and TDE scenarios, but normal CCSNe rapidly inject oxygen-rich ejecta and drive N/O back to lower values. 
Therefore, the observed combination of high N/O ratios and stellar ages requires the oxygen production from normal CCSNe to be strongly suppressed. In our models, this condition is implemented as a high failed-SN fraction among massive stars.}
By incorporating failed supernovae for stars with $M \gtrsim 15\,M_\odot$, the oxygen enrichment is effectively suppressed, allowing our models to maintain high N/O ratios over extended timescales and simultaneously account for the observed stellar ages and [N/O] ratios of seven out of the eight galaxies in our sample.
Recent observations in the Andromeda galaxy have reported a failed supernova from a $13\,M_\odot$ star \citep{2024arXiv241014778D}.
In the high-N/O galaxies, most stars above $15\,M_\odot$ likely undergo failed supernovae.
This explains the absence of a significant increase in oxygen despite high nitrogen abundance.
However, for one object (the blue symbol; GLASS\_150008), our three models fail to simultaneously explain the observed stellar age and [N/O] ratio within their $1\sigma$ uncertainties. 
At approximately $2\times10^{7}$ yr, CCSNe from stars with $M \lesssim 15\,M_\odot$ release a large amount of oxygen, which significantly decreases the N/O ratio.
To explain the ages and N/O ratios, we develop extended models that incorporate the effects of AGB stars and galactic outflows.

The solid curves in the right panel of Figure \ref{fig:NOvsage} illustrate the WR, TDE, and SMS models with the additional contribution from AGB stars. 
To incorporate this effect, we add the ejecta of AGB stars in the mass range of $1\text{--}6\,M_\odot$ \citep{N13papaer} to our baseline models.
We also calculate a model where AGB yields are added to the standard CCSN scenario for stars between $9\text{--}120\,M_\odot$ (magenta curve).
Around $7\times10^{7}$ yr, the onset of AGB enrichment leads to an increase in the N/O ratio; however, the simultaneous oxygen supply from CCSNe ($9\text{--}15\,M_\odot$) keeps the ratio around $\rm [N/O] \sim -0.5$. 
The blue symbol (GLASS\_150008), which could not be explained by our models in the left panel of Figure \ref{fig:NOvsage} due to its older age, remains inconsistent with the model tracks within its $1\sigma$ uncertainty. 
This indicates that the combination of CCSNe and AGB stars alone is insufficient to reproduce the observed high [N/O] ratios in our sample.

To further enhance the [N/O] ratios, we consider a scenario where galactic outflows preferentially remove CCSN ejecta while AGB stars continue to supply nitrogen, as discussed in \cite{2025A&A...697A..96R}. 
The dashed curves in the right panel of Figure \ref{fig:NOvsage} represent these models, where outflows are triggered at the same time as the onset of AGB enrichment ($7\times10^{7}$ yr).
Assuming that 90\% of the accumulated ejecta is expelled, a significant portion of the oxygen is removed from a galaxy.
In this case, the nitrogen from AGB stars becomes the dominant chemical enrichment, allowing the [N/O] ratio to increase up to $\sim 1.5$. 
The blue symbol is now successfully explained by these AGB+outflow models. Although no direct evidence of outflows has been reported for GLASS\_150008 to date, our results suggest that such events might have occurred in its past evolutionary history.
If galactic outflows are sufficiently powerful, the combined CCSN and AGB models can reproduce high [N/O] ratios.
This implies that for the specific case of GLASS\_150008, nitrogen sources such as WR stars, SMSs, or TDEs are not strictly required.
However, the majority of the other galaxies in our sample are significantly younger than the timescale for AGB star enrichment.
Therefore, for these younger systems, the WR, SMS, and TDE scenarios remain essential to account for the observed high nitrogen abundances at such early evolutionary stages.

By incorporating failed supernovae for stars with $M \gtrsim 15\,M_\odot$, along with the effects of AGB stars and galactic outflows, our models can successfully account for the observed N/O ratios and ages of our sample, assuming they represent {\sc Hii} regions formed via a single starburst.
However, a significant gap in the N/O ratio remains between $2\times10^{7}$ and $7\times10^{7}$ yr.
This period corresponds to the interval between the onset of CCSNe from stars with $M \lesssim 15\,M_\odot$ (which enrich the ISM with oxygen) and the subsequent nitrogen injection from AGB stars.
The existence of this gap suggests that our current understanding of the chemical enrichment history in high N/O galaxies may still be incomplete.

We acknowledge the limitations of our one-box chemical evolution models, which assume instantaneous and uniform mixing of stellar yields throughout the ISM.
The simultaneous achievement of high [N/O] and near-solar [Ne/O] ratios could be influenced by spatial inhomogeneities or episodic star-formation activities.
For instance, these abundance ratios might represent different enrichment zones within the galaxy that are not yet fully mixed.
While our current integrated spectra provide limited information on spatial distributions, future IFU observations could offer further insights into whether these abundance ratios arise from localized enrichment processes or a more uniform evolutionary path.

\red{A related systematic uncertainty is the internal ionization and density structure of the nebulae.
If low- and high-ionization lines arise from different gas components, a single-density abundance analysis may not fully describe all of the observed lines.
This issue is particularly relevant for GHZ2, for which \citet{2026OJAp....960281C} show that single-density models cannot reproduce the observed spectrum and that a stratified ISM with both low-/intermediate-density gas and high-density regions ($\log(n_{\rm e}/{\rm cm^{-3}})\gtrsim4$) is required.
More generally, multi-zone effects may be important when combining low- and high-ionization lines \citep{2024A&A...689A..78M}.
For the lensed galaxies in our sample, differential magnification may also change the observed contribution of compact high-ionization regions relative to the rest of the galaxy.
In Section~\ref{sec:abundance}, we isolate the effect of electron density on the line emissivities within our abundance-analysis framework, while keeping the relative contributions of different gas components fixed.
This calculation does not replace full multi-zone photoionization modeling or account for changes in the relative line contributions from different gas components.
Full multi-zone modeling may therefore be required to determine robust absolute oxygen abundances, especially for GHZ2.
Nevertheless, the UV line-ratio abundance estimates used for GHZ2 are insensitive to density over the relevant range, and adopting a higher density for the high-ionization gas in the other galaxies does not remove the nitrogen enhancement.
Thus, density effects alone are unlikely to remove the need for suppressed oxygen production from normal CCSNe.
}

\section{Summary} \label{sec:summary} 
We study the origin of high N/O galaxies observed with JWST at high redshift.
We reduce the raw data of eight high N/O galaxies from JWST/NIRSpec and measure the line fluxes of hydrogen, oxygen, nitrogen, carbon, neon, sulfur, argon, and iron to estimate their abundance ratios through a homogeneous analysis.
We compare the abundance ratios of our galaxies with our chemical evolution models incorporating WR, SMS, and TDE scenarios to investigate the origin of nitrogen enrichment. 
The main results of this paper are summarized below:

\begin{itemize}
    \item We investigate multiple elemental abundance ratios (C/O, Ne/O, S/O, Ar/O, and N/O) in our high-$z$ galaxies. 
    We find that C/O, Ne/O, S/O, and Ar/O are consistent with those of local star-forming galaxies \citep{2006A&A...448..955I, 2019ApJ...874...93B}, while the N/O ratios are significantly elevated, confirming the preferential nitrogen enrichment previously reported. 
    When compared with Mrk 996, a known local WR galaxy, our high-$z$ galaxies show similar abundance ratios. 
    The WR models adopting the wind yields of \cite{2018ApJS..237...13L}, which include both the WN and WC phases, fail to reproduce the observed abundance ratios of Mrk 996, despite the confirmed presence of both WNL and WCE stellar populations in this galaxy. 
    This result indicates that the chemical enrichment pattern observed in Mrk 996 is better explained by the TDE models than by the WR wind models.

    \item The observed combination of elevated N/O, depleted C/O, and near-solar Ne/O ratios cannot be simultaneously reproduced by WR wind enrichment alone, as WR winds do not produce sufficient neon. By incorporating [Ne/C] as a sensitive diagnostic of the CCSN contribution, we find that these abundance patterns are best explained by a dominant WR population with a minor ($\sim$0.1\%) CCSN component. This small CCSN fraction is required to supply the necessary neon without overproducing oxygen, implying that the majority of massive stars in these high-$z$ environments may undergo failed supernovae.
    Incorporating more complex nucleosynthesis scenarios, such as multi-dimensional jet-driven supernova models \citep{2024ApJ...974..310L}, may be required to further improve the fit to the observed abundance patterns.

    \item For the [S/O], [Ar/O], and [Fe/O] ratios, the relevant emission lines are not detected in our high-$z$ galaxies. The large upper limits of these abundance ratios cannot provide strong constraints on the model.
    The Ne/O ratios are useful for distinguishing between models because emission lines are detected in our high-$z$ galaxies.

    \item For the majority of our sample (seven out of eight galaxies), the high [N/O] ratios at young ages ($\lesssim 2\times10^{7}$ yr) are best reproduced by a scenario where most massive stars ($M \gtrsim 15\,M_\odot$) undergo failed supernovae. This process effectively prevents the over-production of oxygen, allowing early nitrogen sources like WR stars, SMSs, or TDEs to dominate the chemical enrichment.

    \item For older systems such as GLASS\_150008 ($\sim 9\times10^{7}$ yr), where early nitrogen sources alone are insufficient, the observed [N/O] ratio and stellar age are explained by the onset of AGB star enrichment combined with powerful galactic outflows. 
    In this scenario, the outflows preferentially expel oxygen-rich CCSN ejecta, ensuring that nitrogen from AGB stars remains the dominant chemical signature even at later stages.

\end{itemize}

\section*{Acknowledgments}
We thank Masayuki. Tanaka, Yuichi. Matsuda, Daichi. Kashino, Hiroya. Umeda, Akinori. Matsumoto, Hajime. Fukushima, Satoshi. Kikuta, Hidenobu. Yajima, and Yi Xu for giving us helpful comments.
This work is based on observations made with the NASA/ESA/CSA James Webb Space Telescope. 
These observations are associated with programs ERS-1324 (GLASS), ERS-1345 (CEERS), GO2478, and GO3073.
The authors acknowledge the GLASS, CEERS, JADES, GO2478, and GO3073 teams led by Tommaso Treu, Steven L. Finkelstein, Daniel Eisenstein, Daniel Stark, and Marco Castellano, respectively, for developing their observing programs with a zero-exclusive access period.
This work is based on observations made with the NASA/ESA/CSA James Webb Space Telescope. The data were obtained from the Mikulski Archive for Space Telescopes at the Space Telescope Science Institute, which is operated by the Association of Universities for Research in Astronomy, Inc., under NASA contract NAS 5-03127 for JWST.
We are grateful to all teams for their extensive efforts in developing and conducting these observation programs.

This publication is based on work supported by the World Premier International Research Center Initiative (WPI Initiative), MEXT, Japan, 
KAKENHI (25H00674) through the Japan Society for the Promotion of Science, and the joint research program of the Institute of Cosmic Ray Research (ICRR), the University of Tokyo.
K. Nomoto has been supported by the World Premier International Research
Center Initiative (WPI), MEXT, Japan, and JSPS KAKENHI grant Nos.
JP23K03452 and JP25K01046.
This work has been supported by JSPS KAKENHI Grant Numbers JP24KJ1160 (K. Watanabe), JP24K07102 (K. Nakajima), JP25KJ0828 (M. Nakane) and JP25K07361 (M. Onodera).

The English in this paper was partially refined with the assistance of ChatGPT (OpenAI 2022).

\bibliography{main}{}
\bibliographystyle{aasjournal}

\end{document}